\documentclass[%
 pra,
 longbibliography,
 superscriptaddress,
 amsmath,amssymb,
reprint,%
nofootinbib
]{revtex4-1}

\usepackage{graphicx}
\usepackage{dcolumn}
\usepackage{bm}

\usepackage{url}
\usepackage{verbatim}
\usepackage{lineno}
\usepackage{aeguill}
\usepackage[latin1]{inputenc}
\usepackage{color}
\usepackage{lineno}
\usepackage{braket}

\draft 

\begin{document}



\title{ Conditional probability and  interferences in generalized measurements with or without definite causal order
} 



\author{M. Trassinelli}
\email[]{martino.trassinelli@insp.jussieu.fr}
\affiliation {Institut des NanoSciences de Paris, CNRS, Sorbonne Université, F-75005 Paris, France}

\date{\today}

\begin{abstract}
In the context of generalized measurement theory, the Gleason-Busch theorem assures the unique form of the associated probability function. 
Recently, in Flatt \text{et al.} Phys. Rev. A \textbf{96}, 062125 (2017), the case of subsequent measurements has been treated, with the derivation of the Lüders rule and its generalization (Kraus update rule). 
Here we investigate the special case of subsequent measurements where an intermediate measurement is a composition of two measurements ($a \text{ or } b$) 
and the case where the causal order is not defined ($a \text{ and } b \text{ or } b \text{ and } a$).
In both cases interference effects can arise.
We show that
the associated probability cannot be written univocally, and the distributive property on its arguments cannot be taken for granted.
The two probability expressions correspond to the Born rule and the classical probability; they are related to the intrinsic possibility of obtaining definite results for the intermediate measurement.
For indefinite causal order, a causal inequality is also deduced.
The frontier between the two cases is investigated in the framework of generalized measurements with a toy model, a Mach-Zehnder interferometer with a movable beam splitter.
\end{abstract}



\maketitle 

%
%
%
%
%
%
%
%
%

\section{Introduction} 

\begin{figure}
\centering
\includegraphics[width=\columnwidth]{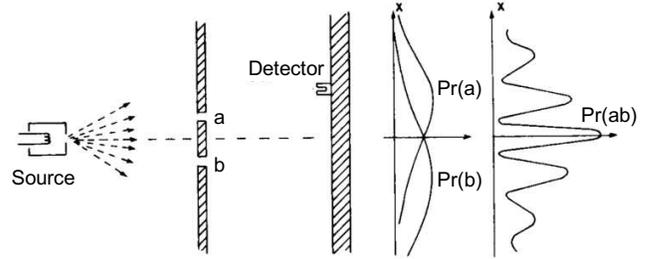}
\caption{Scheme of Young's slits experiment. Adapted from Ref.~\cite{FeynmanLeightonSands}.}
\label{fig:slits}
\end{figure}

In Quantum Mechanics, probabilities are obtained by the squared modulus of complex amplitudes, which give rise to interference phenomena. 
In the common example of Young's setup composed of a source, two slits and a screen or movable detector as represented in Fig.~\ref{fig:slits}, the probability to detect an emitted particle in a position $x$ on the backstop wall  is given by 
\begin{multline}
Pr(ab) = |\psi_a + \psi_b|^2 =\\
= Pr(a) + Pr(b) + 2 \sqrt{Pr(a) Pr(b)} \cos \left( \arg(\psi_a \psi_b^*) \right) , \label{eq:interference}
\end{multline}
where $\psi_a$, $\psi_b$ are the complex probability amplitudes associated with each slit and $Pr(a)=|\psi_a|^2$,  $Pr(b)=|\psi_b|^2$ are the probabilities associated to the opening of the single slits.
The above expression is substantially different from the classical probability sum rule
\begin{equation}
Pr^{C}(ab) = Pr(a) + Pr(b), \label{eq:classic}
\end{equation}
where interference terms are not present.


The probability function for the quantum case is strictly connected to the Hilbert space structure, where systems are described with respect to a defined basis and where the complex numbers mentioned above correspond to coordinates.
With some minimal requirements on the probability function $Pr$, in 1957 Gleason \cite{Gleason1957} demonstrated that $Pr$ is univocally defined in a Hilbert space by the trace rule 
\begin{equation}
Pr(i) = tr(\rho P_i), \label{eq:trace}
\end{equation}
where $\rho=\ket{\varphi}\bra{\varphi}$ is the density matrix of the prepared system and $P_i=\ket{i}\bra{i}$ is the projector on the state of interest. 
In the case of an initial pure state $\ket{\varphi}=\ket{s}$, Eq.~\eqref{eq:trace} corresponds to the Born rule with $Pr(i)= |\braket{i | s} | ^2$.
Gleason's theorem has some limitations; it is valid only for Hilbert spaces with a dimension larger than two and for projective von Neumann measurements \cite{VonNeumann}.

In the framework of the general measurement formalism of positive-operator-valued measures (POVM, also called probability operator measures), in 2003, Busch \cite{Busch2003} extended Gleason's theorem for any dimension and for imperfect measurements described by positive operators, effects $E_i$, instead of projectors.
Recently (2017), in the same context of POVM, Flatt, Barnett and Croke applied the Gleason-Busch theorem to subsequent measurements  \cite{Flatt2017}.
Considering the operators $E_i$ and $F_j$ associated with the measurements $i$ and $j$, with $i$ before $j$, Flatt and coworkers proved that $Pr(i,j)$ takes the general form
\begin{equation}
Pr(i,j) = tr \left( F_j \sum_k K_{ik}^{} \rho K^\dag_{ik} \right), \label{eq:probij}
\end{equation}
where the operators $K_{ik}$ are related to the effects by $E_i = \sum_k K^\dag_{ik} K^{}_{ik}$.
From the above equation and the corresponding one for the conditional probability $Pr(i | j)$, the Kraus update rule \cite{Kraus1971,Kraus} \begin{equation}
\rho  \to \rho'_i = \frac {\sum_k K_{ik}^{} \rho K^\dag_{ik}} {tr \left( \sum_k K^\dag_{ik}  K_{ik}^{} \right)} \label{eq:update}
\end{equation}
for the state update of the system $\rho'_i$ after the measurement $i$ is recovered.
It is worth noting that the Kraus update rule, and its particular case of the Lüders rule \cite{Luders1950} valid for ideal measurement and determined by the von Neumann projection postulate, are derived from first principles.
There is no need of other postulates than the description of states via the Hilbert space and a few basic requirements for the probability function.

In this article, we apply the formalism of subsequent generalized measurements to the 
case with two possible and mutually exclusive intermediate measurements (a two-way interferometer) with indefinite measurement order.
For both cases, interferences can occur.
With the introduction of a new notation of the probability function arguments
with respect to Flatt et al., we will show that  two possible expressions of the final probability $Pr$ can be derived from Eq.~\eqref{eq:trace}.
These two expressions correspond to Eqs.~\eqref{eq:interference} and \eqref{eq:classic} in the example of Young's slits, and are related
to the possibility of distinguishing or not the intermediate measurement.
The difference between the two forms is the order of the arguments of the probability expressions, where the distributive property can not be taken for granted.
The violation of the distributive property in Quantum Mechanics is not new and it has been pointed out since the early years of its formulation \cite{Birkhoff1936,Piron} and extensively discussed in Quantum Logic.
Its connection to the extension of classical probability to quantum probability is well discussed in the literature in the case of perfect projective measurement \cite{BeltramettiCassinelli,Cassinelli1983a,Cassinelli1983b,Hughes,Trassinelli2018}.
For imperfect general measurements, when positive operators are considered instead of projectors, some work has been performed by Busch and collaborators \cite{Busch2006,Busch}.
Here we present a general discussion about the probability function for the distinguishable and indistinguishable path cases (the particle-like and wave-like behaviors) in the case of  imperfect (unsharp) measurements.

For Young's slits,
the frontier between the 
different cases
and the domain of validity of Eqs.~\eqref{eq:interference} and \eqref{eq:classic}, has been extensively discussed in the past. 
Experimentally, it has been explored in the last decades through investigations of interference effects with molecules with larger and larger masses.
Diffraction of large inorganic and organic molecules with masses beyond 25000 atomic mass units has been obtained \cite{Nairz2001,Fein2019,Brand2020}.
Here, we discuss this frontier in the context of generalized measurements considering a Mach-Zehnder interferometer with movable beam splitter.
This toy model, introduced in the past by Haroche et al.  \cite{Bertet2001,Haroche}, has the interesting feature of allowing to pass from one case to the other continuously, simply considering a variation of the mass of the movable beam splitter.

In the case of measurements with indefinite order, extensively discussed in the literature in the last years \cite{Oreshkov2012,Brukner2014,Branciard2015,Castro-Ruiz2018,Zych2019,Wechs2019,Henderson2020}, we will show that they can be treated with the same approach as for the two-path interferometer.
Here too, the presence of interference or not is related in this case to two possible, but not equivalent, expressions of the probability function.


\hfill 

\section{Probability for subsequent measurements}

\subsection{Introduction of new notations}

Taking inspiration from the Quantum Logic approach \cite{BeltramettiCassinelli,Cassinelli1983a,Cassinelli1983b,Hughes,Ballentine1986,DallaChiaraQIL,Trassinelli2018} and the propositional definition of probability \cite{Cox,Fine,Jaynes}, we introduce a new notation with  the logical operators ``$\land$'' and ``$\lor$'' to unambiguously discuss the joint probability of series of subsequent measurements.
The \textit{conjunction} operator ``$\land$'' is equivalent to ``AND'' in normal language and to the comma in the previously introduced notation $Pr(i,j)$.
The \textit{disjunction} operator ``$\lor$'' is equivalent to ``OR'' also indicated with the ``$+$'' operator (in Refs.~\cite{Busch2003,Flatt2017} as example). 
Particular attention has to be payed for measurements $i,j$ that are incompatible.
In this case, the logical operator ``$\land$'' is not well defined \cite{Hughes,Cassinelli1983a,Cassinelli1983b,Ballentine1986}, except if the order of subsequent measurements is defined.
As already pointed out in the \textit{consistent histories} interpretation of Quantum Mechanics \cite{Griffiths1984,Griffiths,Omnes1992}, differently from standard logic,  the operator ``$\land$'' is not symmetric with respect to $i,j$ with $i \land j \ne j \land i$.
With this notation, the joint probability defined above for a measurement $j$ obtained after a measurement $i$ can be written as
\begin{equation}
\wp(j \land i | s) \equiv Pr(i,j),
\end{equation}
where we explicitly indicate the system preparation $s$, which is in fact connected to the possible measurement outcomes. 
We also invert the order of $i,j$ to clearly indicate the sequential order of the measurement or preparation from right to left (preparation $s$, first measurement $i$ and second measurement $j$). 

\subsection{Rewriting probabilities}

\begin{figure}
\centering
\includegraphics[width=0.7\columnwidth]{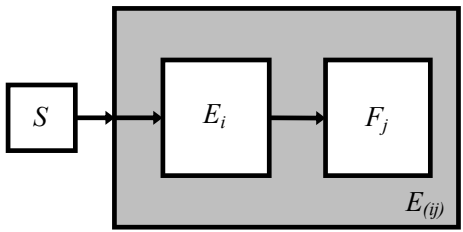}
\caption{Scheme of two subsequent measurements. Adapted from Ref.~\cite{Flatt2017}.}
\label{fig:subsequential}
\end{figure} 

Before treating in detail the Young's slits problem with the new introduced notation, we shall rewrite the properties and assumptions of the probability function used by Flatt et al. \cite{Flatt2017} that lead to Eq.~\eqref{eq:update}.
We consider a set of positive-semidefinite operators (effects) $E_i$ of the same POVM with $\sum_i E_i = I$.
The requirement properties of the probability function $\nu(E_i) = Pr(i)$ for the Gleason-Busch theorem are 
\begin{align*}
\text{(P1)}& \quad 0 \le \nu(E_i) \le 1. \\
\text{(P2)}& \quad \nu(I) = 1.   \\
\text{(P3)}& \quad \nu(E_i + E_j \ldots ) = \nu(E_i) +  \nu(E_j) + \ldots\\
\end{align*}
The function $\nu(E_i)$ is in fact a map from the full set of effects $\mathcal{E(H)}$ acting on the Hilbert space $\mathcal{H}$: $E \to \nu(E)$ with $\nu(E) \in [0,1]$.

With our notation, the previous propositions become
\begin{align*}
\text{(P1*)}& \quad 0 \le \wp(i | s) \le 1. \\
\text{(P2*)}& \quad \wp(\mathcal{I}  | s) = 1.   \\
\text{(P3*)}& \quad \wp(i \lor j \lor \ldots | s) = \wp( i | s) +  \wp(j | s ) + \ldots\\
\end{align*}
where $i,j$ are the measurements that correspond to the effects $E_i,E_j$  and $\mathcal{I}= \bigvee_i i$ measurement correspond to the identity operator $I$.

When two subsequent measurements are considered together, Flatt et al. introduced the new function 
\begin{equation}
\mu_\nu^i(F_j) = \nu(E_{(ij)}) =  Pr(i,j) \label{eq:notation_flatt}
\end{equation}
for the action of the effect $F_j$ after the action of  $E_i$ and  $E_{(ij)}$ indicating the cumulative effect (see Fig.~\ref{fig:subsequential}).
In addition, the following assumptions are considered by Flatt et al.
\begin{align*}
\text{(A1)}& \quad 0 \le \mu_\nu^i(F_j)  \le \nu(E_{i})  < 1. \\
\text{(A2)}& \quad \mu_\nu^i(I) = \nu(E_i).   \\
\text{(A3)}& \quad \mu_\nu^i(F_j + F_k + \ldots) = \mu_\nu^i(F_j) + \mu_\nu^i(F_k) + \ldots. \\
\end{align*}

With our notation, we consider on the same level the measurements $j$ and $i$ and the operator ``$\land$'' indicates the measurement order.
The assumptions (A1--2) can simply be rewritten as
\begin{align*}
\text{(A1*)}& \quad 0 \le \wp(j \land i | s) \le  \wp( i | s) \le 1. \\
\text{(A2*)}& \quad \wp( \mathcal{I} \land i  | s) = \wp(i | s).   \\
\end{align*}
(A2*) is now a tautology.
For (A3), the rewriting is ambiguous.
$\mu_\nu^i(F_j + F_k + \ldots)$ can be written in fact in two different forms:
\begin{equation}
\wp((j \land i) \lor (k \land i)  | s) \label{eq:P_dist}
\end{equation} 
or 
\begin{equation}
\wp((j \lor k) \land i | s). \label{eq:P_indist}
\end{equation}

Another situation that cannot be easily treated with the formalism of Flatt et al. is the case of subsequent measurements with indefinite causal order.
Equation \eqref{eq:notation_flatt} implies in fact a defined causal order because of the nesting of the two probability functions.
The treatment of indefinite causal order phenomena and the associated possible interferences is well defined  for elementary processes in the context of quantum field theory (e.g. in the relativistic Compton scattering \cite{Itzykson-Zuber}).
When a subsequent interaction with measuring detectors is considered, the situation is more complicated and it has been the center of interest of several works in the last years \cite{Oreshkov2012,Brukner2014,Branciard2015,Henderson2020}.
Considering two measurements represented by two effects $E_i$ and $E_j$ with an indefinite order and a final measurement represented by $F_d$, similarly to Eqs.~\eqref{eq:P_dist} and \eqref{eq:P_indist}, the associated probability can be written in two ways
\begin{equation}
\wp(k \land [(i \land j) \lor (j \land i) ]| s) \label{eq:P_dist_c}
\end{equation} 
or 
\begin{equation}
\wp( (k \land i \land j) \lor (k \land j \land i)  | s). \label{eq:P_indist_c}
\end{equation}

In the following sections, it will be shown that the formulas \eqref{eq:P_dist}--\eqref{eq:P_indist} and \eqref{eq:P_dist_c}--\eqref{eq:P_indist_c} will lead to different probability expressions.
These different expressions, obtained from a unique definition of the probability function, will be the key point of the present work.
%

 \subsection{General considerations for POVM operators}
 
To investigate the difference between Eqs.~\eqref{eq:P_dist} and \eqref{eq:P_indist}, we come back the specific example of Young's slits where we consider the possibility to measure or flag the passage through each slit. 
Before that, a short introduction to generalized measurements is mandatory.
In the framework of POVM, the single measurements are described by the positive-valued operators $E_\ell=K_\ell^\dag K_\ell^{}$, where
$K_\ell$ operators are determined by the unitary interaction between the system we want to study and the detector, both considered  as quantum systems.
The general expression for $K_\ell$ is given by \cite{Barnett,Laloe}
\begin{equation}
K_\ell   = \sum_{i,j} \alpha_{ij} \braket{\ell^{det} | \Phi^{det}_i}\ket{j}\bra{i},
\end{equation}
where $\alpha_{ij}$ depend on the action of the unitary matrix $U_{int}$ describing the interaction between the system and the detector.
The initial state is described by $\ket{\Psi_i^0} = \ket{i} \ket{\Phi_0}$, where $\ket{i}$ and  $\ket{\Phi_0}$ describe the initial state of the system and the detector, respectively.
After their mutual interaction, the system and detector states are described by $\ket{\Psi_i^0} \to \ket{\Psi_i} = \ket{\varphi_i}  \ket{\Phi^{det}_i} = \sum_j \alpha_{ij} \ket{j} \ket{\Phi^{det}_i}$.
$\ket{\Phi^{det}_i}$ describes the detector state after the interaction with the system in an initial state $\ket{i}$.
Finally, $\ket{\ell^{det}}$ represents the detector state corresponding to the macroscopic outcome of the measurement device.
In the case of a non-destructive measurement, the above formula is simplified to
\begin{equation}
K_\ell =  \sum_{i} \braket{\ell^{det} | \Phi^{det}_i}\ket{i}\bra{i}. \label{eq:Kgen}
\end{equation}

\begin{figure}
\centering
\includegraphics[width=0.7\columnwidth]{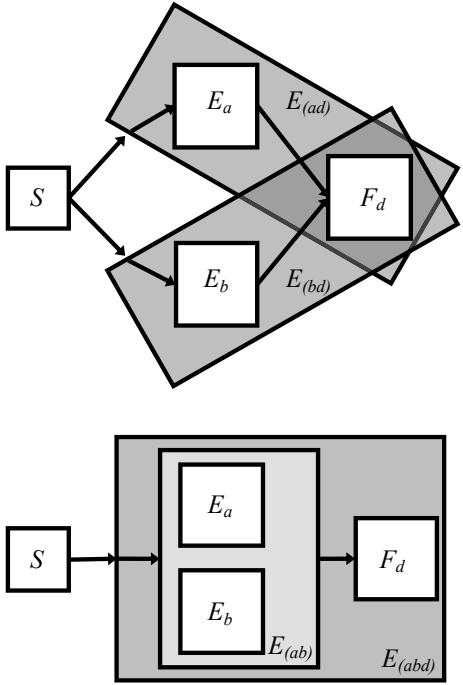}
\caption{Schemes of subsequential measurements corresponding to the Young's slits experiment for the case where the path of the particle can be detected (top) or not (bottom).}
\label{fig:dist-indist}
\end{figure}


\section{ Interferences in a two-path interferometer}

\subsection{Distinguishable paths}  
In the case of Young's slits, we consider that the detection of the path taken by the particle is possible and is non-destructive.
The formula corresponding to Eq.~\eqref{eq:P_dist} becomes $\wp((d \land a) \lor (d \land b) | s)$ and depends on the operators $K_a, K_b$ and $K_d$.
 $K_a, K_b$ are related to the detection of the path $a$ or $b$, and the corresponding effects are $E_a = K_a^\dag K_a^{}, \ E_b = K_b^\dag K_b^{}$.
$K_d$ is related to the detection $d$ on the screen with $F_d = K_d^\dag K_d^{}$
The combination of $E_a$ and $E_b$ with $F_d$ can be assimilated to the effects $E_{(ad)}$ and  $E_{(bd)}$ as in Eq.~\eqref{eq:notation_flatt}, and for which the property (P3)/(P3*) can be applied.
In this case we have 
\begin{multline}
\wp((d \land a) \lor (d \land b) | s) = \wp(d \land a | s) + \wp(d \land b | s) = \\
 = tr ( F_d K_a^{} \rho K_a^\dag) + tr(F_d K_b^{} \rho K_b^\dag).  \label{eq:KaKb}
\end{multline}

The above equation corresponds to the classic probability sum rule, i.e. the particle-like probability in Eq.~\eqref{eq:classic}.
The fact that we can decompose the measurement in two separate operators $E_{(ad)}$ and  $E_{(bd)}$ (Fig.~\ref{fig:dist-indist}, top) implicitly means that the different paths can be distinguished and we have just a duplicated version of the basic subsequent measurement represented in Fig.~\ref{fig:subsequential}.
This case can be easily treated with the formalism introduced by Flatt et al. with the introduction of the probability functions $\mu_\nu^a(F_d)$ and $\mu_\nu^b(F_d)$.

In the case of ideal projective measurements, we have $E_a = K_a = P_a = \ket{a}\bra{a}$ and $E_b = K_b = P_b = \ket{b}\bra{b}$ where we used the properties of projectors $P_i^\dag = P_i$  and $P_i P_i = P_i$.
The above equation then becomes \cite{Trassinelli2018}
\begin{multline}
 \wp((d \land a) \lor (d \land b) | s) = \\
  = | \braket{d | U_{(ad)} | a}\braket{a | U_{(sa)} | s} |^2  + |\braket{d | U_{(bd)} | b}\braket{b | U_{(sb)} | s} |^2, \label{eq:AorB_ideal}
\end{multline}
where the unitary operators $U$ correspond to the evolution of the different parts of the apparatus.

The expression  of $\wp((d \land a) \lor (d \land b) | s)$ can also be directly obtained by the trace reduction of the density matrix $\rho$ with respect to detector base $\ket{a^{det}}$ and $\ket{b^{det}}$.
In this case we have
\begin{equation}
\wp((d \land a) \lor (d \land b) | s) =   tr(F_d  \rho_r),  \label{eq:KaORb_reduced}
\end{equation}
with $\rho_r = tr_{a^{det},b^{det}} \left( \ket{\Psi_i} \bra{\Psi_i} \right)$ and where  $\ket{\Psi_i} = \sum_j \alpha_{ij} \ket{j} \ket{\Phi_i}$.
From the linearity of the trace operator, it is easy to verify that the previous expression is equivalent to Eq.~\eqref{eq:KaKb}.
This indicates that the use of the trace over the undetected $\ket{a^{det}}, \ket{b^{det}}$ states implicitly implies an interaction between the system and the which-path detectors, even if they are not directly involved in the measurement.

\subsection{Indistinguishable paths and discussion} \label{sec:indist_slits}

In the case we can not distinguish which path is taken by the particle, the $a \lor b$ cannot be decomposed and we have to deal with the expression
\begin{equation}
\wp(d \land (a \lor b) | s) =  tr( F_d K_{a \lor b}^{} \rho K_{a \lor b}^\dag). \label{eq:KaORb}
\end{equation}
The operator $K_{a \lor b}$ can be built in three different ways:
\begin{enumerate}
\item from path-detectors with the same final state after the interaction with the system ($\ket{a^{det}}=\ket{b^{det}}$),
\item from a complementary measurement $c$ (e.g. a series of detectors on the slit walls),
\item via a detector state $\ket{q^{det}}$ belonging to the span generated by the vectors $\ket{a^{det}}$ and $\ket{b^{det}}$.
\end{enumerate}
As we will see, a genuine $a \lor b$ measurement is related to the first two cases only.
The third
approach is in fact related to the \textit{quantum eraser} case and it is discussed separately in the next section.

In the case where the measurement of the passage of a particle in one path or the other induces the same detector state $\ket{ab^{det}} \equiv \ket{a^{det}} = \ket{b^{det}}$, the operator corresponding to $a \lor b$ can be written as
\begin{equation}
K_{a \lor b} = \sum_{i=a,b} \braket{\ell^{det} | ab^{det}}\ket{i}\bra{i} =  \sum_{i=a,b} K_i, \label{eq:KaORb_no-distinction}
\end{equation}
where $K_a$ and $K_b$ are associated to the effects $E_a$ and $E_b$ represented in the bottom of Fig.~\ref{fig:dist-indist} and $\ket{\ell^{det}}$ is a generic state corresponding to the $a \lor b$ measure ($\ket{\ell^{det}} = \ket{ab^{det}}$ for an ideal measurement).
Note that this is not the case for Kraus operators $K_{ik}$ as in Eq.~\eqref{eq:update} where different detector states $ \Phi^{det}_k$ correspond to the same system state $\ket{i}$.
Here in opposite, different system states $\ket{a}$ and $\ket{b}$ correspond to the same detector state $\ket{ab^{det}}$, and the trace operator in  Eqs.~\eqref{eq:update} and \eqref{eq:KaORb} cannot be separated into different terms.

If we consider a complementary measurement $c$ to both $a$ and $b$ measurements, we have  that $c \land a =0, c \land b = 0$ and $c = \mathcal{I} - a \lor b$.
$E_{a \lor b}$ corresponds to the absence of signal in the measurement $E_c$, then, using the property of the set of effects of the POVM for which $\sum_{i=a,b,c} E_i = I$, we have $E_{a \lor b} = I - E_c = E_a + E_b$.
$K_{a \lor b}$ can be written as \cite{Barnett,Busch,AulettaQM}
\begin{equation}
K_{a \lor b} = U_{a \lor b} \sqrt{E_a + E_b},   \label{eq:KaORb_complementary}
\end{equation}  where $U_{a \lor b}$ is a unitary matrix that depends on the details of the interaction between $\ket{a^{det},b^{det}}$  and the propagating particle-wave.


In the case of ideal projective measurements for both situations discussed above, $K_{a \lor b}$ can be explicitly written.
In this case we have that $E_{a \lor b} = E_a + E_b = P_a +P_b$ and  Eq.~\eqref{eq:KaORb} becomes  \cite{Trassinelli2018} (see also Refs.~\cite{BeltramettiCassinelli,Cassinelli1983b,Busch})
\begin{multline}
 \wp(d \land (a \lor  b) | s) = \\
 = | \braket{d | U_{(ad)} | a}\braket{a | U_{(sa)} | s}   + \braket{d | U_{(bd)} | b}\braket{b | U_{(sb)} | s} |^2, \label{eq:AB_ideal}
\end{multline}
which is equivalent to the quantum form of the probability in Eq.~\eqref{eq:interference}, i.e. equivalent to the Born rule.

In the general case represented in Eqs.~\eqref{eq:KaKb} and  \eqref{eq:KaORb} (and in the particular case in Eqs.~\eqref{eq:AorB_ideal} and \eqref{eq:AB_ideal}), 
\begin{equation}
\wp(d \land (a \lor  b) | s) \ne \wp((d \land a) \lor (d \land b) | s)
\end{equation}
and the distributive property on the arguments of $\wp$ is violated.
Equations \eqref{eq:KaKb} and \eqref{eq:AorB_ideal} reproduce the sum rule valid for the classical probability $Pr^C$ (Eq.~\eqref{eq:classic}). 
However, Eqs.~\eqref{eq:KaORb} and \eqref{eq:AB_ideal} present additional interference terms and are compatible with the Born rule.
If the distributive property is considered valid, the two expressions 
should be equivalent.
But the validity of distributivity cannot in fact be taken for granted.
As anticipated in the introduction, the violation of the distributive law in quantum phenomena is well known since the early years of Quantum Mechanics \cite{Birkhoff1936,Putnam1969}.
In particular in Quantum Logic \cite{Piron1972,Piron,BeltramettiCassinelli,Hughes,Auletta,DallaChiara,DallaChiaraQIL} this is related to the properties of orthomodular lattices, associated to sets of yes/no experiments, where the distributivity on their elements is not always valid.

For the indistinguishable case, the measurement $a \lor b$ corresponds to an atomic operator $E_{a \lor b}\equiv E_{(ab)}$ that cannot be decomposed in terms of $E_a,E_b$.
The cumulative effect $E_{(abd)}$ depends then on all three measurements $a$, $b$ and $d$ and can be represented by the scheme in the bottom of Fig.~\ref{fig:dist-indist}.

\subsection{The quantum eraser revisited} \label{sec:qe}
In the Quantum Logic context, a measurement representing $a \lor b$ can be built from a vector $\ket{q^{det}} = \alpha  \ket{a^{det}}+ \beta \ket{b^{det}}$ \cite{Hughes}, with $\alpha,\beta \ne 0$, which belongs to the span generated by the vectors $\ket{a^{det}}$ and $\ket{b^{det}}$.
Using Eq.~\eqref{eq:Kgen} with $\ell = q$, we can then write
\begin{equation}
K_{a \lor b} \equiv K_q = \alpha^* K_a + \beta^* K_b, \label{eq:KaORb_bis}
\end{equation}
where $|\alpha|^2+|\beta|^2 = 1$ for a normalized probability.
Once inserted in Eq.~\eqref{eq:KaORb}, the above expression gives rise to mixed $\braket{a^{det} | b^{det}}$ terms and then to interference phenomena.
This is in fact the case of the \textit{quantum eraser} \cite{Scully1991,Herzog1995,Kim2000,Busch2006,Weisz2014,Ma2016}, where instead of the direct path detection via $\ket{a^{det}}, \ket{b^{det}}$, a combination of them is considered and interference terms appear.

This is a situation not equivalent to the case with a complementary  measurement $c = \mathcal{I} - a \lor b$.
Even if we recover the presence of interferences with the use of $\ket{q^{det}}$ instead of $\ket{a^{det}}$ or $\ket{b^{det}}$, we are dealing with a single measurement $q$ that corresponds to the probability $\wp ( d \land q | s)$, and not $\wp ( d \land (a \lor b) | s)$.
Similarly to $a,b$ measurements, we could consider the alternative measurement given by the vector $\ket{r^{det}} = - e^{i \phi}\beta \ket{a^{det}} + e^{i \phi} \alpha \ket{b^{det}}$ orthogonal to $\ket{q^{det}}$.
When both possible measurements $q$ and $r$ are considered, we can write down the probabilities $\wp ( (d \land q)  \lor (d \land r)) | s)$ and $\wp ( d \land (q \lor r) | s)$.
$\ket{r^{det}},\ket{q^{det}}$ and $\ket{a^{det}},\ket{b^{det}}$ are two different bases describing the detection and they are related by a unitary transformation.
Because of the property of the unitary transformation, it can be demonstrated (see App.~\ref{app:QE} for the detailed calculations)
that the combination of the two measurements $r$ and $q$ and  the which-path $a$ and $b$ are completely equivalent and
\begin{equation}
\wp ( (d \land q)  \lor (d \land r)) | s) = \wp ( (d \land a)  \lor (d \land b)) | s).
\end{equation}
The interference terms present in the separate terms  $\wp ( d \land q | s)$ and $\wp ( d \land r | s)$, completely compensate in $\wp ( (d \land q)  \lor (d \land r)) | s)= \wp ( d \land q | s) + \wp ( d \land r | s)$  like in the well known results on the quantum eraser.

For the case of $\wp ( d \land (q \lor r)) | s)$ probability, the situation is more complicated because it depends on the values of $\alpha,\beta,\phi$ but also on the choice of $\alpha', \beta'$ for building $K_{q \lor r} = \alpha' K_q + \beta' K_r$.
With this last consideration, we can conclude that in fact the construction of $K_{a \lor b}$ via Eq.~\eqref{eq:KaORb_bis} is not equivalent to a genuine which-path ignorance, but it is a special case where a different detector state basis is considered.


\subsection{A toy model with a Mach-Zehnder interferometer}

The fundamental difference between distinguishable and indistinguishable cases, i.e., the use of Eq.~\eqref{eq:KaORb_reduced} or  Eq.~\eqref{eq:KaORb_complementary} for the probability function, is the coupling between the considered system and the possible which-path detector(s) and/or the environment, but also the information that can be extracted from the detector(s) outputs.
Such a coupling has been extensively studied in the context of decoherence theory \cite{Zurek2003,Schlosshauer2005}.
In this section we consider a very simple case to investigate the limits of Eqs.~\eqref{eq:KaORb_reduced} and  \eqref{eq:KaORb_complementary} in terms of effects thanks to a toy system where we can continuously tune the detectability of the taken path. 

\begin{figure}
\centering
\includegraphics[width=0.8\columnwidth]{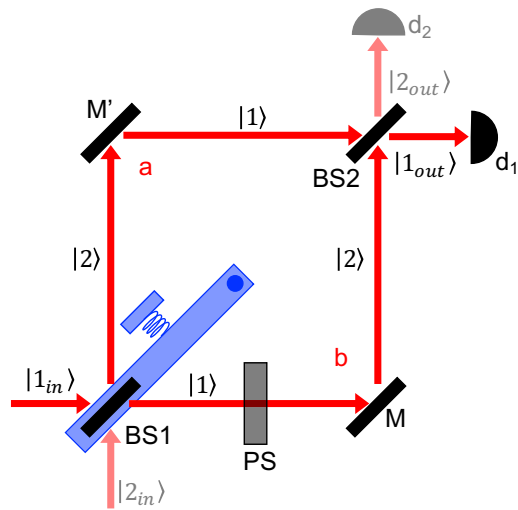}
\caption{Scheme of the Mach-Zehnder interferometer with a movable beam splitter BS1, a phase shifter with $\phi$ with two possible incoming beams $\ket{1_{in}}$ and  $\ket{2_{in}}$ and outputs $\ket{1_{out}}$ and  $\ket{2_{out}}$ measured by the detectors $d_1$ and $d_2$.
Photons parallel to the incoming photon (horizontal propagation in the figure) are indicated by the states $\ket{1}$ and with $\ket{2}$ otherwise (vertical propagation).}
\label{fig:MZI}
\end{figure}

We consider a Mach-Zehnder interferometer with a movable beam splitter (represented in Fig.~\ref{fig:MZI}), an example discussed in the literature and realized experimentally with atoms in resonant cavities \cite{Bertet2001,Haroche}.
Here we treat the problem in terms of effects in a POVM framework.
A discussion of the Mach-Zehnder interferometer in terms of unsharp detection has been already discussed by Busch and Shilladay \cite{Busch2006}.
In this past work, the unsharpness of the detection was studied in terms of measurement mixing between the two paths, like in the quantum eraser case discussed in Sec.~\ref{sec:qe}.
The cases of  distinguishability or indistinguishability of the paths were also treated, but not the frontier between them, which on the contrary is the main subject of the following paragraphs.
A more general treatment of the continuous passage between distinguishable and indistinguishable cases has been done in the past by B.-G. Englert \cite{Englert1996}.
The visibility of interferences and of the \textit{distinguishability} quantity (in terms of the amount of which-way information can be extracted) is discussed, but not the relation to the probability function in the context of POVM.

The system considered here is composed by a single-photon source emitting monochromatic photons $\ket{1_{in}}$ interacting with: a movable beam splitter $BS1$, two mirrors $M, M'$, a phase shifter $PS$ that induces a phase $\phi$, a second (fixed) beam splitter $BS2$ and two detectors $d_1$ and $d_2$, following the scheme represented in Fig.~\ref{fig:MZI}.
The movable beam splitter, with a mass $m$, can move with respect to a pivot and is connected to a fixed part by a spring that corresponds to a resonant angular frequency $\omega$.
The beam splitter-spring system is described by a harmonic oscillator with energy spectrum $\mathcal{E}_n = \omega \hbar (n + \frac 1 2)$.
When the photon is reflected from the first beam splitter, a momentum kick $\Delta P = \sqrt{2} p$, with $p= 2 \pi \hbar / \lambda$ the impulse of the photon, 
is transferred to $BS1$ with a translation from its ground state $\ket{0}_{BS}$ to the coherent state $\ket{\alpha_{BS}}$ with $\alpha = ip / \sqrt{m \omega \hbar}$ \cite{Bertet2001,Haroche}.

In analogy to the Young's slits, we can consider the interferometer arm with the reflection from the movable beam splitter as the path $a$, and path $b$ otherwise (see Fig.~\ref{fig:MZI}).


In the case of a fixed beam splitter, the state of the beam splitter itself does not change after the passage of the photon and the state corresponding to the photon is
\begin{multline}
\ket{1_{in}} \to \ket{\phi} = \\
- \frac 1 2 \ket{1_{out}} - \frac i 2 \ket{2_{out}} - \frac {e^{i \phi}} 2 \ket{1_{out}} + \frac {i e^{i \phi}} 2 \ket{2_{out}} 
\label{eq:bs_fixed}
\end{multline}
The probability of detecting something on the detector $d_1$ depends on the operator $K_{d_1} = \ket{0_{out}} \bra{1_{out}}$ and the corresponding effect $F_{d_1} = K_{d_1}^\dag K^{}_{d_1} = \ket{1_{out}}\bra{1_{out}}$.
Because of the impossibility of determining the path taken by the photon, the corresponding probability is $\wp(d_1 \land (a \lor b)|s)$. 
The complementary detection $c$ representing $K_{a \lor b}$ could be constitued by a series of detectors around the beam splitter $BS1$, like the wall detection in the case of the Young's slits, to insure the interaction (reflection or transmission) of the incoming photon $\ket{1_{in}}$ with $BS1$.
The probability is then given by
\begin{equation}
\wp(d_1 \land (a \lor b)|s) = tr(F_{d_1} \rho') = \frac 1 2 \left[ 1 +  \cos (\phi ) \right], \label{eq:MZI_AB}
\end{equation}
with $\rho' = \ket{\phi}\bra{\phi}$ and $\ket{\phi}$ given by Eq.~\eqref{eq:bs_fixed}.
We recover the standard formula of the Mach-Zehnder interferometer \cite{Busch2006,AulettaQM}.


We consider now that the beam splitter $BS1$ can move and that its state after the recoil is described by the coherent state $\ket{\alpha_{BS}}$.
Considering the initial state $\ket{1_{in}} \ket{0_{BS}}$ describing the photon-beam splitter system, after the interaction between the incoming photon with the first beamsplitter $BS1$ (and the mirrors $M$ and $M'$ and the second beam splitter $BS2$), the photon/mirror state $\ket{\phi}$ is described by 
\begin{multline}
\ket{1_{in}} \ket{0_{BS}} \to = \ket{\phi} \ket{\Phi_{BS}} = \\
= - \frac 1 2 \ket{1_{out}}\ket{\alpha_{BS}} - \frac i 2 \ket{2_{out}}\ket{\alpha_{BS}} +\\
- \frac {e^{i \phi}} 2 \ket{1_{out}}\ket{0_{BS}} + \frac {i e^{i \phi}} 2 \ket{2_{out}}\ket{0_{BS}} ,
\end{multline}
where $\ket{\phi}$ is state of the photon at the exit of the interferometer and $\ket{\Phi_{BS}}$ is the state of the movable beam splitter after the passage of the photon.

The operator $K_b = \braket{0_{BS} |  \Phi_{BS} }  \ket{\phi} \bra{1_{in}}$ can be associated to the branch $b$ where there is no momentum transfer to $BS1$, which remains in the $\ket{0_{BS}}$ state.
For the branch $a$, we cannot directly use $\braket{\alpha_{BS} | \Phi_{BS} }  \ket{\phi} \bra{1_{in}}$ as $K_a$ operator.
Due to the non-orthogonality of $\ket{\alpha_{BS}}$ and $\ket{0_{BS}}$, this leads to the possibility of having $E_a + E_b >1$, violating the basic POVM properties. 
Considering that we can identify a coherent state only if its corresponding signal is above the quantum shot noise of the system, instead of $\ket{\alpha_{BS}}$ we can consider its Gram-Schmidt orthogonalization $\ket{\alpha_{BS}'}$ with respect to $\ket{0_{BS}}$
\begin{equation}
\ket{\alpha_{BS}'} = \frac {\ket{\alpha_{BS}} - \braket{0_{BS} | \alpha_{BS}}  \ket{0_{BS}} }{\sqrt{1-|\braket{0_{BS} | \alpha_{BS}}|^2}}.
\end{equation}
The corresponding which-path operators are then
\begin{multline}
K_a = \braket{\alpha_{BS}' | \phi } \bra{1_{in}}  = \frac 1 2  \sqrt{1- |\braket{0_{BS} | \alpha_{BS}}|^2} \ket{1_{out}} \bra{1_{in}} +\\
            - \frac i 2 \sqrt{1- |\braket{0_{BS} | \alpha_{BS}}|^2}\braket{ \alpha_{BS}' | \alpha_{BS}} \ket{2_{out}} \bra{1_{in}} 
\end{multline}
and
\begin{multline}
K_b = \braket{0_{BS} | \Psi } \bra{1_{in}} = \\
= - \frac 1 2 \braket{0_{BS} | \alpha_{BS}} \ket{1_{out}} \bra{1_{in}} - \braket{0_{BS} | \alpha_{BS}} \frac i 2 \ket{2_{out}} \bra{1_{in}} + \\
- \frac {e^{i \phi}} 2  \ket{1_{out}} \bra{1_{in}} +  \frac { i e^{i \phi}} 2\ket{2_{out}} \bra{1_{in}}. 
\end{multline}

The corresponding probabilities of the single paths become
\begin{align}
&\wp (d_1 \land a | s) =  tr(F_{d_1} K^{}_a \rho K^\dag_a) = \frac 1 4 \left( 1 -  | \braket{0_{BS} | \alpha_{BS}} |^2 \right) \\
&\wp (d_1 \land b | s) =  tr(F_{d_1} K^{}_b \rho K^\dag_b) = \frac 1 4 \left|  1 + e^{i \phi} \braket{0_{BS} | \alpha_{BS}} \right|^2 =  \\ \nonumber 
& \ =  \frac 1 4 \left[ 1 + | \braket{0_{BS} | \alpha_{BS}} |^2 + 2 \Re e (e^{i \phi} \braket{0_{BS} | \alpha_{BS}} ) \right] 
\end{align}
Finally, we have then
\begin{equation}
\wp ((d_1 \land a) \lor (d_1 \land b) | s) = \frac 1 2 \left[ 1 + e^{- \frac {| \alpha |^2} 2} \cos (\phi ) \right],  \label{eq:MZI_AorB}
\end{equation}
where we used $\braket{\alpha_{BS} | 0_{BS}}= e^{- \frac {| \alpha |^2} 2}$.

\begin{figure}
\centering
\includegraphics[clip,trim = 0 10  0 10,width=0.8\columnwidth]{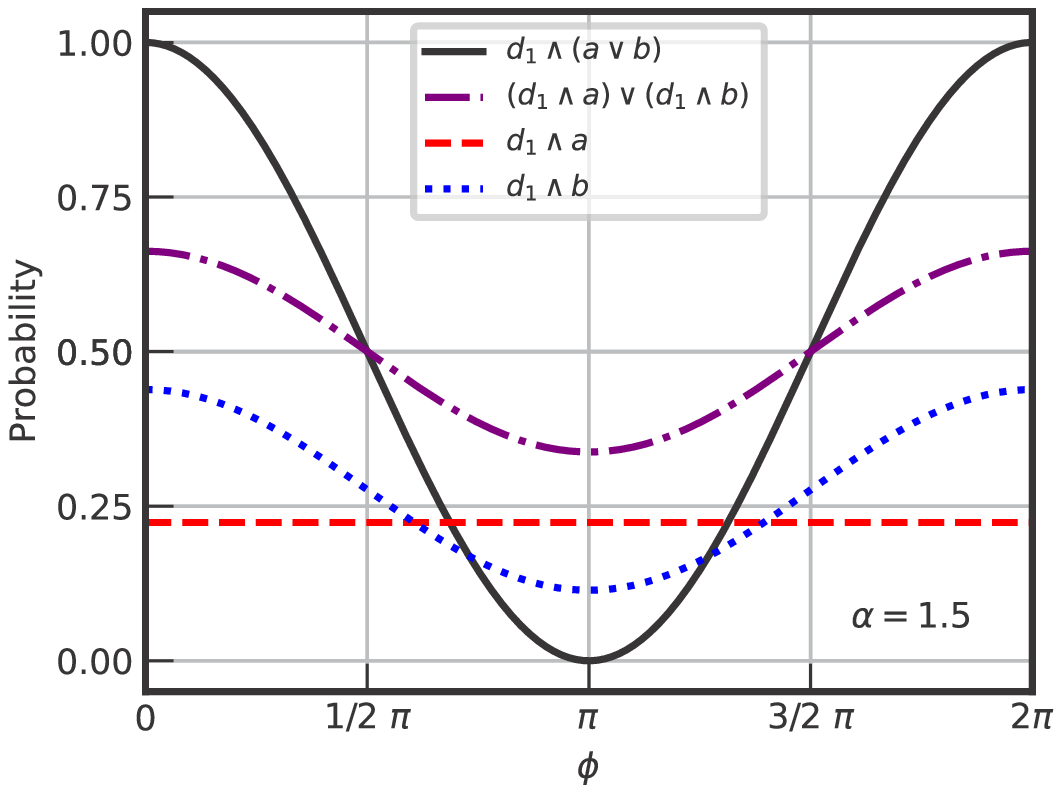}\\
\includegraphics[clip,trim = 0 10  0 10,width=0.8\columnwidth]{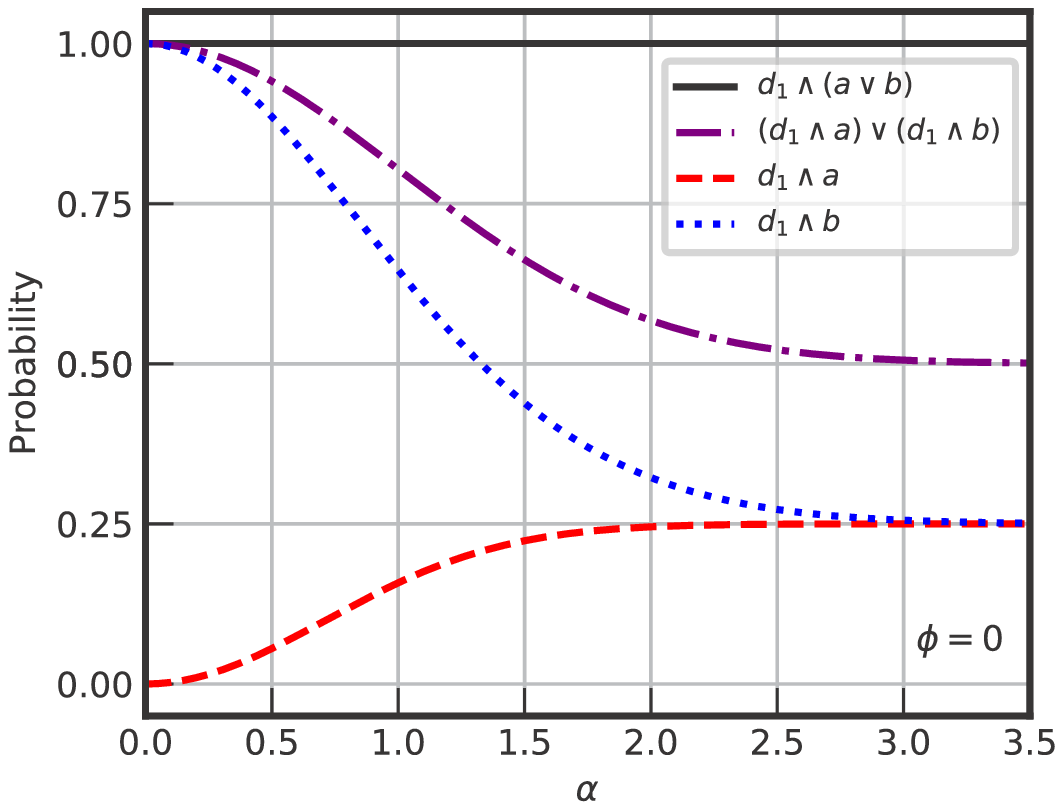}
\caption{Dependency of the different probability expressions on $\phi$ and $\alpha$ with the fixed value $\alpha=1.5$ (top)
and with the fixed value $\phi=0$ (bottom). }
\label{fig:scans}
\end{figure} 

As we can see in Fig.~\ref{fig:scans}, for each probability relative to a specific path, an interference term is always present and is proportional to the overlap between the $\ket{0}_{BS}$ and $\ket{\alpha_{BS}}$ states. 
Only the probability corresponding to the path $b$ is sensitive to the phase of the $PS$.

In the limit $\braket{0_{BS} | \alpha_{BS}} \to 0$ (corresponding to $\alpha \to \infty$ and $m \to 0$, see  Fig.~\ref{fig:scans} bottom), we have a pure particle-like behavior with $\wp (d_1 \land a | s) = \wp (d_1 \land b | s) = 1/ 4 $ and $\wp ((d_1 \land a) \lor (d_1 \land b) | s) = 1/2$.

In the limit $\braket{0_{BS} | \alpha_{BS}} \to 1$ (corresponding to $\alpha \to 0$ and $m \to \infty$, see  Fig.~\ref{fig:scans} bottom),  we have instead $\wp (d_1 \land a | s) = 0$ and $\wp (d_1 \land b | s) = 1$. 
From the detection of $\ket{0}_{BS}$, no information on the taken path can be extracted. 
This is similar to the case 3 discussed in Sec.~\ref{sec:indist_slits}, where both path detectors provide the same output.
In this limit case $\wp (d_1 \land b | s)$ (and then $\wp ((d_1 \land a) \lor (d_1 \land b) | s)$) is \textit{de facto} equivalent to $\wp (d_1 \land (a \lor  b) | s)$ treated in the previous section.
The behavior of the different formulas as function of $\phi$ and $\alpha$ is shown in Fig.~\ref{fig:scans}.

Except to the limit case with $\braket{0_{BS} | \alpha_{BS}} \to 0$ ($m \to 0$), the two equations \eqref{eq:MZI_AB} and \eqref{eq:MZI_AorB} lead to different forms of the probability function.
Once more, the expressions $d_1 \land (a \lor b)$ and $(d_1 \land a) \lor (d_1 \land b)$ cannot be considered equivalent with the violation of the distributivity property.


\section{Interferences in subsequent measurements with indefinite causal order}

\subsection{General considerations for indefinite causal order measurements}

Another two-way system where interference effects occur is the case of subsequent measurements with an indefinite causal order.
To treat such class of subsequent measurements, the new formalism of the process matrices has been developed  in the last years \cite{Oreshkov2012,Brukner2014,Henderson2020}, and the presence of interference effects in measurements with indefinite order has been proven experimentally \cite{Rubino2017}. 
In particular it has been demonstrated that, for two subsequent measurements $a$ and $b$, the probability relation \cite{Oreshkov2012,Brukner2014,Branciard2015}
\begin{equation}
Pr(a,b) = \lambda Pr^{a \preccurlyeq b} (a,b) + (1- \lambda) Pr^{b \preccurlyeq a} (a,b),  \label{eq:causal_equality}
\end{equation}
valid for a probabilistic composition between two measurements with defined causal order (where the measurement order is indicated by the operator ``$\preccurlyeq$'' and with $0 \le \lambda  \le 1$) is violated for indefinite causal phenomena.

With the probability definition and notation introduced in the previous sections, both cases of two-way interferences and two causal orders are simply discussed on the same footing and without introducing new notations in the probability function like $Pr^{a \preccurlyeq b}$.

In the case a single causal sequence of emissions from a source $s$, with a measurement $a$ followed by a measurement $b$ and final detection $d$, the corresponding probability is 
\begin{equation}
\wp(d \land b \land a | s) = tr(F_d K^{}_b K^{}_a \rho K^\dag_a K^\dag_b),
\end{equation}
where  $K_a$ and $K_b$ are the Kraus operators associated to the effects $E_a$ and $E_b$ corresponding to the intermediate measurements, respectively, and $F_d$ is the effect corresponding to the final detection.
Note that due to the implicit prior of the order ``$a$ after $b$'' in the notation $Pr^{a \preccurlyeq b}$, in our notation $Pr^{a \preccurlyeq b}(a,b)$ corresponds in fact to 
$\wp(d | b \land a \land s) = \wp(d \land b \land a | s) /  \wp(b \land a | s)$. 

Similarly to the which-path case, if we consider the same subsequents measurements but where the order between $a$ and $b$ is unknown,  the probabilities $\wp((d \land b \land a) \lor (d \land a \land b)) | s$ and $\wp (d \land [(b \land a) \lor ( a \land b) ]| s)]$ lead to two different final expressions.

\subsection{Distinguishable causal order}
This case corresponds to the situation where we can implicitly determine the measurement order, similarly to the case where each path in the Young's slits can be determined, and the associated probability writing is $\wp((d \land b \land a) \lor (d \land a \land b)) | s$.
The different sequences of measurements can be associated to the effects $E_{(abd)}$ and  $E_{(bad)}$ and each corresponding probability can be constructed from a nested version of Eq.~\eqref{eq:notation_flatt}.
Moreover, in this case the property (P3)/(P3*) can be applied:
\begin{multline}
\wp((d \land b \land a) \lor (d \land a \land b)| s) = \wp(d \land a \land b | s) + \wp(d \land b \land a | s) = \\
= tr(F_d K^{}_b K^{}_a \rho K^\dag_a K^\dag_b) + tr(F_d K^{}_a K^{}_b \rho K^\dag_b K^\dag_a) = \\
= \wp(a \land b  | s) \wp(d | a \land b \land  s) +  \wp(b \land a  | s) \wp(d | b \land a \land  s). \label{eq:causal_ordered}
\end{multline}
The last equality of the above equation allows to appreciate the equivalence to Eq.~\eqref{eq:causal_equality}, which is manifestly not violated where $\wp(a \land b  | s) = \lambda$.

\subsection{Indistinguishable causal order and discussion}

If the order of the measurement is intrinsically indefinite, $\wp (d \land [(b \land a) \lor ( a \land b)] | s)]$ has to be considered with
\begin{equation}
\wp(d \land [(b \land a) \lor ( a \land b)])| s) = tr(F_d K^{}_{a \overline{\land} b} \rho K^\dag_{a \overline{\land} b}), \label{eq:causal_noordered}
\end{equation}
where the Kraus operator $K^{}_{a \overline{\land} b} $ corresponds to the indefinite ordered measurement $(b \land a) \lor ( a \land b)$.
Compared to Eq.~\eqref{eq:causal_ordered}, additional interference terms can be present and we have
\begin{equation}
\wp(d \land [(b \land a) \lor ( a \land b)])| s) \neq \wp((d \land b \land a) \lor (d \land a \land b)| s).
\end{equation}

In order to appreciate the differences between the two expressions, we can consider ideal measurements for $a$ and $b$ with the assumption that they can be described by non-commutative and non-orthogonal projectors $P_a$ and $P_b$ acting in the large Hilbert space composed by the inputs and outputs of the two measurements \cite{Flatt2017,Oreshkov2012}.
With these assumptions, Eq.~\eqref{eq:causal_ordered} can be written as
\begin{multline}
\wp((d \land b \land a) \lor (d \land a \land b)| s)  = \\
= tr(P_d P_b P_a \rho P_a P_b) + tr(P_d P_a P_b \rho P_b P_a). \label{eq:causal}
\end{multline}
On the other hand, Eq.~\eqref{eq:causal_noordered} becomes
\begin{multline}
\wp(d \land [(b \land a) \lor ( a \land b)]| s)  = \\
= tr(P_d P_b P_a \rho P_a P_b) + tr(P_d P_a P_b \rho P_b P_a) + \\
+ tr(P_d P_b P_a \rho P_b P_a) + tr(P_d P_b P_a \rho P_b P_a) + \ldots
\end{multline}
where the cross terms $ tr(P_d P_b P_a \rho P_b P_a)$ and $tr(P_d P_b P_a \rho P_b P_a)$ are present (plus additional terms related to the non-commutativity between  $P_a$ and $P_b$).
The above equation violates Eq.~\eqref{eq:causal_equality}, as obtained in the work of Brukner and coworkers with the process matrices formalism \cite{Oreshkov2012,Brukner2014,Branciard2015}.

\section{Conclusion}

In conclusion, we present a formulation of the probability function in the context of generalized measurements for subsequent detections with several possible paths, and for subsequent detections with indefinite order.
From the assumption of the Hilbert space structure for the description of systems, Gleason-Busch theorem assures that the trace of the density operator univocally defines the form of the probability function. 
Flatt and coworkers demonstrate that from this result, when subsequent measurements are considered, the Kraus updating rule is reconstructed.
Here we apply the same methodology to a two-path case with a renewed notation but also to the case of intermediate measurements without define causal order.

For the two-path case, two different expressions of the probability are found, $\wp( (d \land a) \lor (d \land b) | s)$ and $\wp(d \land ( a \lor b) | s)$, which are related to the possibility of distinguishing or not the trajectory in the measurement system.
In fact, the distributive property of the probability function arguments cannot be taken for granted.
From the first expression, the classical law or probability $Pr^C(a \lor b) = Pr^C(a) + Pr^C(b)$ is recovered.
The use of the reduced trace over the undetected states of the path-detectors leads to this same expression.
With regards to the $\wp(d \land ( a \lor b) | s)$, the associated operator $K_{a \lor b}$ to the $ a \lor b$ measurement, can be interpreted ambiguously.
$K_{a \lor b}$ can be built from measurements corresponding to the same final detector state independently from the path, or from a complementary measurement of $a$ and $b$ ($c = NOT ( a \lor b)$).
Both approaches lead
to a final expression corresponding to the standard Born rule for the case of perfect projective measurements.
If $K_{a \lor b}$ is constructed by a mixing of  path-detector states, we recover the situation of the \textit{quantum eraser}.
We are in fact considering the probability $\wp( d \land q | s)$ associated with the state $\ket{q^{det}} = \alpha \ket{a^{det}} + \beta \ket{b^{det}}$, which depends on the choice of constants $\alpha$ and $\beta$, i.e. a special case of which-path ignorance.

The frontier between the intrinsic possibility to distinguish a path or not is related to the coupling of the studied system with the path-detectors and/or the environment.
This topic is widely studied in the literature, and in particular in the context of decoherence theory.
Here we consider the very simple case of a  Mach-Zehnder interferometer with a movable beam splitter, which is also well known in the literature but treated here in the context of generalized measurements.
We demonstrate here that by varying the mass of the beam splitter, we can continuously pass from the distinguishable path case, where $\wp( (d \land a) \lor (d \land b) | s)$ is valid, to the  indistinguishable path case, where $\wp(d \land ( a \lor b) | s)$ should be used instead.
This toy model reveals once more the complementarity of nature, but also underlines once more the advantages of generalized measurement theory with respect to ideal projective measurements, where unsharp detections revealing particle-like and wave-like behavior at the same time can be treated unambiguously. 

With the same probability notation introduced here, another two-way problem is treated, namely the sequence of detections with an indefinite causal order.
For the case of two indefinite ordered measurements $a$ and $b$, two different formulation of the probability function are allowed, $\wp((d \land b \land a) \lor (d \land a \land b)| s)$ and $\wp(d \land [(b \land a) \lor ( a \land b)])| s)$.
Similarly to the two-path case, the difference between the two expressions is related to the possibility to distinguish or not the causal order of the measurements.
The first expression corresponds to a statistical sum of probabilities relative to the two possible measurement sequences with definite order. 
In this case the ``causal equality'' (Eq.~\eqref{eq:causal_equality}) is valid.
In the second expression, additional interference terms appear with the violation of the causal equality, similarly to the results obtained with the process matrices formalism.

\begin{acknowledgments}
I would like to thank very much M.~Romanelli and M. Walschaers for their constructive critics to the previous versions of the manuscript, but also C. Fabre, S. Reynaud, V. Parigi and N. Paul for the stimulating discussions and support. 
I would like also to thank A. Caticha and N. Carrara for the encouragement and suggestions after a first talk on a primordial version of the presented work.
\end{acknowledgments}

\bibliography{interference_POVM_new}

\begin{thebibliography}{54}%
\makeatletter
\providecommand \@ifxundefined [1]{%
 \@ifx{#1\undefined}
}%
\providecommand \@ifnum [1]{%
 \ifnum #1\expandafter \@firstoftwo
 \else \expandafter \@secondoftwo
 \fi
}%
\providecommand \@ifx [1]{%
 \ifx #1\expandafter \@firstoftwo
 \else \expandafter \@secondoftwo
 \fi
}%
\providecommand \natexlab [1]{#1}%
\providecommand \enquote  [1]{``#1''}%
\providecommand \bibnamefont  [1]{#1}%
\providecommand \bibfnamefont [1]{#1}%
\providecommand \citenamefont [1]{#1}%
\providecommand \href@noop [0]{\@secondoftwo}%
\providecommand \href [0]{\begingroup \@sanitize@url \@href}%
\providecommand \@href[1]{\@@startlink{#1}\@@href}%
\providecommand \@@href[1]{\endgroup#1\@@endlink}%
\providecommand \@sanitize@url [0]{\catcode `\\12\catcode `\$12\catcode
  `\&12\catcode `\#12\catcode `\^12\catcode `\_12\catcode `\%12\relax}%
\providecommand \@@startlink[1]{}%
\providecommand \@@endlink[0]{}%
\providecommand \url  [0]{\begingroup\@sanitize@url \@url }%
\providecommand \@url [1]{\endgroup\@href {#1}{\urlprefix }}%
\providecommand \urlprefix  [0]{URL }%
\providecommand \Eprint [0]{\href }%
\providecommand \doibase [0]{http://dx.doi.org/}%
\providecommand \selectlanguage [0]{\@gobble}%
\providecommand \bibinfo  [0]{\@secondoftwo}%
\providecommand \bibfield  [0]{\@secondoftwo}%
\providecommand \translation [1]{[#1]}%
\providecommand \BibitemOpen [0]{}%
\providecommand \bibitemStop [0]{}%
\providecommand \bibitemNoStop [0]{.\EOS\space}%
\providecommand \EOS [0]{\spacefactor3000\relax}%
\providecommand \BibitemShut  [1]{\csname bibitem#1\endcsname}%
\let\auto@bib@innerbib\@empty
\bibitem [{\citenamefont {Feynman}\ \emph {et~al.}(1963)\citenamefont
  {Feynman}, \citenamefont {Leighton},\ and\ \citenamefont
  {Sands}}]{FeynmanLeightonSands}%
  \BibitemOpen
  \bibfield  {author} {\bibinfo {author} {\bibfnamefont {R.P.}\ \bibnamefont
  {Feynman}}, \bibinfo {author} {\bibfnamefont {R.B.}\ \bibnamefont
  {Leighton}}, \ and\ \bibinfo {author} {\bibfnamefont {M.L.}\ \bibnamefont
  {Sands}},\ }\href@noop {} {\emph {\bibinfo {title} {The Feynman Lectures on
  Physics}}}\ (\bibinfo  {publisher} {Pearson/Addison-Wesley},\ \bibinfo {year}
  {1963})\BibitemShut {NoStop}%
\bibitem [{\citenamefont {Gleason}(1957)}]{Gleason1957}%
  \BibitemOpen
  \bibfield  {author} {\bibinfo {author} {\bibfnamefont {A.~M.}\ \bibnamefont
  {Gleason}},\ }\bibfield  {title} {\enquote {\bibinfo {title} {Measures on the
  closed subspaces of a hilbert space},}\ }\href@noop {} {\bibfield  {journal}
  {\bibinfo  {journal} {J. Math. Mech.}\ }\textbf {\bibinfo {volume} {6}},\
  \bibinfo {pages} {885--893} (\bibinfo {year} {1957})}\BibitemShut {NoStop}%
\bibitem [{\citenamefont {{Von Neumann}}(1955)}]{VonNeumann}%
  \BibitemOpen
  \bibfield  {author} {\bibinfo {author} {\bibfnamefont {J.}~\bibnamefont {{Von
  Neumann}}},\ }\href@noop {} {\emph {\bibinfo {title} {Mathematical
  Foundations of Quantum Mechanics}}}\ (\bibinfo  {publisher} {Princeton
  University Press},\ \bibinfo {year} {1955})\BibitemShut {NoStop}%
\bibitem [{\citenamefont {Busch}(2003)}]{Busch2003}%
  \BibitemOpen
  \bibfield  {author} {\bibinfo {author} {\bibfnamefont {P.}~\bibnamefont
  {Busch}},\ }\bibfield  {title} {\enquote {\bibinfo {title} {Quantum states
  and generalized observables: A simple proof of {G}leason's theorem},}\
  }\href@noop {} {\bibfield  {journal} {\bibinfo  {journal} {Phys. Rev. Lett.}\
  }\textbf {\bibinfo {volume} {91}},\ \bibinfo {pages} {120403} (\bibinfo
  {year} {2003})}\BibitemShut {NoStop}%
\bibitem [{\citenamefont {Flatt}\ \emph {et~al.}(2017)\citenamefont {Flatt},
  \citenamefont {Barnett},\ and\ \citenamefont {Croke}}]{Flatt2017}%
  \BibitemOpen
  \bibfield  {author} {\bibinfo {author} {\bibfnamefont {Kieran}\ \bibnamefont
  {Flatt}}, \bibinfo {author} {\bibfnamefont {Stephen~M.}\ \bibnamefont
  {Barnett}}, \ and\ \bibinfo {author} {\bibfnamefont {Sarah}\ \bibnamefont
  {Croke}},\ }\bibfield  {title} {\enquote {\bibinfo {title} {Gleason-busch
  theorem for sequential measurements},}\ }\href@noop {} {\bibfield  {journal}
  {\bibinfo  {journal} {Phys. Rev. A}\ }\textbf {\bibinfo {volume} {96}},\
  \bibinfo {pages} {062125} (\bibinfo {year} {2017})}\BibitemShut {NoStop}%
\bibitem [{\citenamefont {Kraus}(1971)}]{Kraus1971}%
  \BibitemOpen
  \bibfield  {author} {\bibinfo {author} {\bibfnamefont {K.}~\bibnamefont
  {Kraus}},\ }\bibfield  {title} {\enquote {\bibinfo {title} {General state
  changes in quantum theory},}\ }\href {\doibase
  https://doi.org/10.1016/0003-4916(71)90108-4} {\bibfield  {journal} {\bibinfo
   {journal} {Ann. Phys.}\ }\textbf {\bibinfo {volume} {64}},\ \bibinfo {pages}
  {311--335} (\bibinfo {year} {1971})}\BibitemShut {NoStop}%
\bibitem [{\citenamefont {Kraus}\ \emph {et~al.}(1983)\citenamefont {Kraus},
  \citenamefont {Böhm},\ and\ \citenamefont {Dollard}}]{Kraus}%
  \BibitemOpen
  \bibfield  {author} {\bibinfo {author} {\bibfnamefont {K.}~\bibnamefont
  {Kraus}}, \bibinfo {author} {\bibfnamefont {A.}~\bibnamefont {Böhm}}, \ and\
  \bibinfo {author} {\bibfnamefont {J.D.}\ \bibnamefont {Dollard}},\
  }\href@noop {} {\emph {\bibinfo {title} {States, Effects, and Operations
  Fundamental Notions of Quantum Theory}}}\ (\bibinfo  {publisher} {Springer},\
  \bibinfo {year} {1983})\BibitemShut {NoStop}%
\bibitem [{\citenamefont {Lüders}(1950)}]{Luders1950}%
  \BibitemOpen
  \bibfield  {author} {\bibinfo {author} {\bibfnamefont {Gerhart}\ \bibnamefont
  {Lüders}},\ }\bibfield  {title} {\enquote {\bibinfo {title} {Über die
  zustandsänderung durch den meßprozeß},}\ }\href {\doibase
  doi:10.1002/andp.19504430510} {\bibfield  {journal} {\bibinfo  {journal}
  {Ann. Phys.}\ }\textbf {\bibinfo {volume} {443}},\ \bibinfo {pages}
  {322--328} (\bibinfo {year} {1950})}\BibitemShut {NoStop}%
\bibitem [{\citenamefont {Birkhoff}\ and\ \citenamefont {{Von
  Neumann}}(1936)}]{Birkhoff1936}%
  \BibitemOpen
  \bibfield  {author} {\bibinfo {author} {\bibfnamefont {Garrett}\ \bibnamefont
  {Birkhoff}}\ and\ \bibinfo {author} {\bibfnamefont {John}\ \bibnamefont {{Von
  Neumann}}},\ }\bibfield  {title} {\enquote {\bibinfo {title} {The logic of
  quantum mechanics},}\ }\href@noop {} {\bibfield  {journal} {\bibinfo
  {journal} {Ann. Math.}\ }\textbf {\bibinfo {volume} {37}},\ \bibinfo {pages}
  {823--843} (\bibinfo {year} {1936})}\BibitemShut {NoStop}%
\bibitem [{\citenamefont {Piron}(1976)}]{Piron}%
  \BibitemOpen
  \bibfield  {author} {\bibinfo {author} {\bibfnamefont {C.}~\bibnamefont
  {Piron}},\ }\href@noop {} {\emph {\bibinfo {title} {Foundations of quantum
  physics}}}\ (\bibinfo  {publisher} {Benjamin-Cummings Publishing Company},\
  \bibinfo {year} {1976})\BibitemShut {NoStop}%
\bibitem [{\citenamefont {Beltrametti}\ and\ \citenamefont
  {Cassinelli}(1984)}]{BeltramettiCassinelli}%
  \BibitemOpen
  \bibfield  {author} {\bibinfo {author} {\bibfnamefont {E.G.}\ \bibnamefont
  {Beltrametti}}\ and\ \bibinfo {author} {\bibfnamefont {G.}~\bibnamefont
  {Cassinelli}},\ }\href@noop {} {\emph {\bibinfo {title} {The Logic of Quantum
  Mechanics}}}\ (\bibinfo  {publisher} {Cambridge University Press},\ \bibinfo
  {year} {1984})\BibitemShut {NoStop}%
\bibitem [{\citenamefont {Cassinelli}\ and\ \citenamefont
  {Zanghì}(1983)}]{Cassinelli1983a}%
  \BibitemOpen
  \bibfield  {author} {\bibinfo {author} {\bibfnamefont {G.}~\bibnamefont
  {Cassinelli}}\ and\ \bibinfo {author} {\bibfnamefont {N.}~\bibnamefont
  {Zanghì}},\ }\bibfield  {title} {\enquote {\bibinfo {title} {Conditional
  probabilities in quantum mechanics. i.-conditioning with respect to a single
  event},}\ }\href {\doibase 10.1007/BF02721792} {\bibfield  {journal}
  {\bibinfo  {journal} {Nuovo Cim. B}\ }\textbf {\bibinfo {volume} {73}},\
  \bibinfo {pages} {237--245} (\bibinfo {year} {1983})}\BibitemShut {NoStop}%
\bibitem [{\citenamefont {Cassinelli}\ and\ \citenamefont
  {Zanghí}(1984)}]{Cassinelli1983b}%
  \BibitemOpen
  \bibfield  {author} {\bibinfo {author} {\bibfnamefont {G.}~\bibnamefont
  {Cassinelli}}\ and\ \bibinfo {author} {\bibfnamefont {N.}~\bibnamefont
  {Zanghí}},\ }\bibfield  {title} {\enquote {\bibinfo {title} {Conditional
  probabilities in quantum mechanics. ii. additive conditional
  probabilities},}\ }\href {\doibase 10.1007/BF02748966} {\bibfield  {journal}
  {\bibinfo  {journal} {Nuovo Cim. B}\ }\textbf {\bibinfo {volume} {79}},\
  \bibinfo {pages} {141--154} (\bibinfo {year} {1984})}\BibitemShut {NoStop}%
\bibitem [{\citenamefont {Hughes}(1989)}]{Hughes}%
  \BibitemOpen
  \bibfield  {author} {\bibinfo {author} {\bibfnamefont {R.I.G.}\ \bibnamefont
  {Hughes}},\ }\href@noop {} {\emph {\bibinfo {title} {The Structure and
  Interpretation of Quantum Mechanics}}}\ (\bibinfo  {publisher} {Harvard
  University Press},\ \bibinfo {year} {1989})\BibitemShut {NoStop}%
\bibitem [{\citenamefont {Trassinelli}(2018)}]{Trassinelli2018}%
  \BibitemOpen
  \bibfield  {author} {\bibinfo {author} {\bibfnamefont {M.}~\bibnamefont
  {Trassinelli}},\ }\bibfield  {title} {\enquote {\bibinfo {title} {Relational
  quantum mechanics and probability},}\ }\href@noop {} {\bibfield  {journal}
  {\bibinfo  {journal} {Found. Phys.}\ }\textbf {\bibinfo {volume} {48}},\
  \bibinfo {pages} {1092--1111} (\bibinfo {year} {2018})}\BibitemShut {NoStop}%
\bibitem [{\citenamefont {Busch}\ and\ \citenamefont
  {Shilladay}(2006)}]{Busch2006}%
  \BibitemOpen
  \bibfield  {author} {\bibinfo {author} {\bibfnamefont {Paul}\ \bibnamefont
  {Busch}}\ and\ \bibinfo {author} {\bibfnamefont {Christopher}\ \bibnamefont
  {Shilladay}},\ }\bibfield  {title} {\enquote {\bibinfo {title}
  {Complementarity and uncertainty in mach-zehnder interferometry and
  beyond},}\ }\href@noop {} {\bibfield  {journal} {\bibinfo  {journal} {Phys.
  Rep.}\ }\textbf {\bibinfo {volume} {435}},\ \bibinfo {pages} {1--31}
  (\bibinfo {year} {2006})}\BibitemShut {NoStop}%
\bibitem [{\citenamefont {Busch}\ \emph {et~al.}(2016)\citenamefont {Busch},
  \citenamefont {Lahti}, \citenamefont {Pellonpää},\ and\ \citenamefont
  {Ylinen}}]{Busch}%
  \BibitemOpen
  \bibfield  {author} {\bibinfo {author} {\bibfnamefont {P.}~\bibnamefont
  {Busch}}, \bibinfo {author} {\bibfnamefont {P.}~\bibnamefont {Lahti}},
  \bibinfo {author} {\bibfnamefont {J.P.}\ \bibnamefont {Pellonpää}}, \ and\
  \bibinfo {author} {\bibfnamefont {K.}~\bibnamefont {Ylinen}},\ }\href@noop {}
  {\emph {\bibinfo {title} {Quantum Measurement}}}\ (\bibinfo  {publisher}
  {Springer International Publishing},\ \bibinfo {year} {2016})\BibitemShut
  {NoStop}%
\bibitem [{\citenamefont {Nairz}\ \emph {et~al.}(2001)\citenamefont {Nairz},
  \citenamefont {Brezger}, \citenamefont {Arndt},\ and\ \citenamefont
  {Zeilinger}}]{Nairz2001}%
  \BibitemOpen
  \bibfield  {author} {\bibinfo {author} {\bibfnamefont {Olaf}\ \bibnamefont
  {Nairz}}, \bibinfo {author} {\bibfnamefont {Björn}\ \bibnamefont {Brezger}},
  \bibinfo {author} {\bibfnamefont {Markus}\ \bibnamefont {Arndt}}, \ and\
  \bibinfo {author} {\bibfnamefont {Anton}\ \bibnamefont {Zeilinger}},\
  }\bibfield  {title} {\enquote {\bibinfo {title} {Diffraction of complex
  molecules by structures made of light},}\ }\href {\doibase
  10.1103/PhysRevLett.87.160401} {\bibfield  {journal} {\bibinfo  {journal}
  {Phys. Rev. Lett.}\ }\textbf {\bibinfo {volume} {87}},\ \bibinfo {pages}
  {160401} (\bibinfo {year} {2001})}\BibitemShut {NoStop}%
\bibitem [{\citenamefont {Fein}\ \emph {et~al.}(2019)\citenamefont {Fein},
  \citenamefont {Geyer}, \citenamefont {Zwick}, \citenamefont {Kialka},
  \citenamefont {Pedalino}, \citenamefont {Mayor}, \citenamefont {Gerlich},\
  and\ \citenamefont {Arndt}}]{Fein2019}%
  \BibitemOpen
  \bibfield  {author} {\bibinfo {author} {\bibfnamefont {Yaakov~Y.}\
  \bibnamefont {Fein}}, \bibinfo {author} {\bibfnamefont {Philipp}\
  \bibnamefont {Geyer}}, \bibinfo {author} {\bibfnamefont {Patrick}\
  \bibnamefont {Zwick}}, \bibinfo {author} {\bibfnamefont {Filip}\ \bibnamefont
  {Kialka}}, \bibinfo {author} {\bibfnamefont {Sebastian}\ \bibnamefont
  {Pedalino}}, \bibinfo {author} {\bibfnamefont {Marcel}\ \bibnamefont
  {Mayor}}, \bibinfo {author} {\bibfnamefont {Stefan}\ \bibnamefont {Gerlich}},
  \ and\ \bibinfo {author} {\bibfnamefont {Markus}\ \bibnamefont {Arndt}},\
  }\bibfield  {title} {\enquote {\bibinfo {title} {Quantum superposition of
  molecules beyond 25 {kDa}},}\ }\href {\doibase 10.1038/s41567-019-0663-9}
  {\bibfield  {journal} {\bibinfo  {journal} {Nat. Phys.}\ }\textbf {\bibinfo
  {volume} {15}},\ \bibinfo {pages} {1242--1245} (\bibinfo {year}
  {2019})}\BibitemShut {NoStop}%
\bibitem [{\citenamefont {Brand}\ \emph {et~al.}(2020)\citenamefont {Brand},
  \citenamefont {Kialka}, \citenamefont {Troyer}, \citenamefont {Knobloch},
  \citenamefont {Simonovic}, \citenamefont {Stickler}, \citenamefont
  {Hornberger},\ and\ \citenamefont {Arndt}}]{Brand2020}%
  \BibitemOpen
  \bibfield  {author} {\bibinfo {author} {\bibfnamefont {Christian}\
  \bibnamefont {Brand}}, \bibinfo {author} {\bibfnamefont {Filip}\ \bibnamefont
  {Kialka}}, \bibinfo {author} {\bibfnamefont {Stephan}\ \bibnamefont
  {Troyer}}, \bibinfo {author} {\bibfnamefont {Christian}\ \bibnamefont
  {Knobloch}}, \bibinfo {author} {\bibfnamefont {Ksenija}\ \bibnamefont
  {Simonovic}}, \bibinfo {author} {\bibfnamefont {Benjamin~A.}\ \bibnamefont
  {Stickler}}, \bibinfo {author} {\bibfnamefont {Klaus}\ \bibnamefont
  {Hornberger}}, \ and\ \bibinfo {author} {\bibfnamefont {Markus}\ \bibnamefont
  {Arndt}},\ }\bibfield  {title} {\enquote {\bibinfo {title} {Bragg diffraction
  of large organic molecules},}\ }\href {\doibase
  10.1103/PhysRevLett.125.033604} {\bibfield  {journal} {\bibinfo  {journal}
  {Phys. Rev. Lett.}\ }\textbf {\bibinfo {volume} {125}},\ \bibinfo {pages}
  {033604} (\bibinfo {year} {2020})}\BibitemShut {NoStop}%
\bibitem [{\citenamefont {Bertet}\ \emph {et~al.}(2001)\citenamefont {Bertet},
  \citenamefont {Osnaghi}, \citenamefont {Rauschenbeutel}, \citenamefont
  {Nogues}, \citenamefont {Auffeves}, \citenamefont {Brune}, \citenamefont
  {Raimond},\ and\ \citenamefont {Haroche}}]{Bertet2001}%
  \BibitemOpen
  \bibfield  {author} {\bibinfo {author} {\bibfnamefont {P.}~\bibnamefont
  {Bertet}}, \bibinfo {author} {\bibfnamefont {S.}~\bibnamefont {Osnaghi}},
  \bibinfo {author} {\bibfnamefont {A.}~\bibnamefont {Rauschenbeutel}},
  \bibinfo {author} {\bibfnamefont {G.}~\bibnamefont {Nogues}}, \bibinfo
  {author} {\bibfnamefont {A.}~\bibnamefont {Auffeves}}, \bibinfo {author}
  {\bibfnamefont {M.}~\bibnamefont {Brune}}, \bibinfo {author} {\bibfnamefont
  {J.~M.}\ \bibnamefont {Raimond}}, \ and\ \bibinfo {author} {\bibfnamefont
  {S.}~\bibnamefont {Haroche}},\ }\bibfield  {title} {\enquote {\bibinfo
  {title} {A complementarity experiment with an interferometer at the
  quantum-classical boundary},}\ }\href {\doibase 10.1038/35075517} {\bibfield
  {journal} {\bibinfo  {journal} {Nature}\ }\textbf {\bibinfo {volume} {411}},\
  \bibinfo {pages} {166--170} (\bibinfo {year} {2001})}\BibitemShut {NoStop}%
\bibitem [{\citenamefont {Haroche}\ \emph {et~al.}(2006)\citenamefont
  {Haroche}, \citenamefont {Raimond},\ and\ \citenamefont {Press}}]{Haroche}%
  \BibitemOpen
  \bibfield  {author} {\bibinfo {author} {\bibfnamefont {S.}~\bibnamefont
  {Haroche}}, \bibinfo {author} {\bibfnamefont {J.M.}\ \bibnamefont {Raimond}},
  \ and\ \bibinfo {author} {\bibfnamefont {Oxford~University}\ \bibnamefont
  {Press}},\ }\href@noop {} {\emph {\bibinfo {title} {Exploring the Quantum:
  Atoms, Cavities, and Photons}}}\ (\bibinfo  {publisher} {Oxford University
  Press, Oxford},\ \bibinfo {year} {2006})\BibitemShut {NoStop}%
\bibitem [{\citenamefont {Oreshkov}\ \emph {et~al.}(2012)\citenamefont
  {Oreshkov}, \citenamefont {Costa},\ and\ \citenamefont
  {Brukner}}]{Oreshkov2012}%
  \BibitemOpen
  \bibfield  {author} {\bibinfo {author} {\bibfnamefont {Ognyan}\ \bibnamefont
  {Oreshkov}}, \bibinfo {author} {\bibfnamefont {Fabio}\ \bibnamefont {Costa}},
  \ and\ \bibinfo {author} {\bibfnamefont {Caslav}\ \bibnamefont {Brukner}},\
  }\bibfield  {title} {\enquote {\bibinfo {title} {Quantum correlations with no
  causal order},}\ }\href {\doibase 10.1038/ncomms2076} {\bibfield  {journal}
  {\bibinfo  {journal} {Nat. Commun.}\ }\textbf {\bibinfo {volume} {3}},\
  \bibinfo {pages} {1092} (\bibinfo {year} {2012})}\BibitemShut {NoStop}%
\bibitem [{\citenamefont {Brukner}(2014)}]{Brukner2014}%
  \BibitemOpen
  \bibfield  {author} {\bibinfo {author} {\bibfnamefont {Caslav}\ \bibnamefont
  {Brukner}},\ }\bibfield  {title} {\enquote {\bibinfo {title} {Quantum
  causality},}\ }\href {\doibase 10.1038/nphys2930} {\ \textbf {\bibinfo
  {volume} {10}},\ \bibinfo {pages} {259--263} (\bibinfo {year}
  {2014})}\BibitemShut {NoStop}%
\bibitem [{\citenamefont {Branciard}\ \emph {et~al.}(2015)\citenamefont
  {Branciard}, \citenamefont {Araújo}, \citenamefont {Feix}, \citenamefont
  {Costa},\ and\ \citenamefont {Brukner}}]{Branciard2015}%
  \BibitemOpen
  \bibfield  {author} {\bibinfo {author} {\bibfnamefont {Cyril}\ \bibnamefont
  {Branciard}}, \bibinfo {author} {\bibfnamefont {Mateus}\ \bibnamefont
  {Araújo}}, \bibinfo {author} {\bibfnamefont {Adrien}\ \bibnamefont {Feix}},
  \bibinfo {author} {\bibfnamefont {Fabio}\ \bibnamefont {Costa}}, \ and\
  \bibinfo {author} {\bibfnamefont {Caslav}\ \bibnamefont {Brukner}},\
  }\bibfield  {title} {\enquote {\bibinfo {title} {The simplest causal
  inequalities and their violation},}\ }\href {\doibase
  10.1088/1367-2630/18/1/013008} {\bibfield  {journal} {\bibinfo  {journal}
  {New J. Phys.}\ }\textbf {\bibinfo {volume} {18}},\ \bibinfo {pages} {013008}
  (\bibinfo {year} {2015})}\BibitemShut {NoStop}%
\bibitem [{\citenamefont {Castro-Ruiz}\ \emph {et~al.}(2018)\citenamefont
  {Castro-Ruiz}, \citenamefont {Giacomini},\ and\ \citenamefont
  {Brukner}}]{Castro-Ruiz2018}%
  \BibitemOpen
  \bibfield  {author} {\bibinfo {author} {\bibfnamefont {Esteban}\ \bibnamefont
  {Castro-Ruiz}}, \bibinfo {author} {\bibfnamefont {Flaminia}\ \bibnamefont
  {Giacomini}}, \ and\ \bibinfo {author} {\bibfnamefont {Caslav}\ \bibnamefont
  {Brukner}},\ }\bibfield  {title} {\enquote {\bibinfo {title} {Dynamics of
  quantum causal structures},}\ }\href {\doibase 10.1103/PhysRevX.8.011047}
  {\bibfield  {journal} {\bibinfo  {journal} {Phys. Rev. X}\ }\textbf {\bibinfo
  {volume} {8}},\ \bibinfo {pages} {011047} (\bibinfo {year}
  {2018})}\BibitemShut {NoStop}%
\bibitem [{\citenamefont {Zych}\ \emph {et~al.}(2019)\citenamefont {Zych},
  \citenamefont {Costa}, \citenamefont {Pikovski},\ and\ \citenamefont
  {Brukner}}]{Zych2019}%
  \BibitemOpen
  \bibfield  {author} {\bibinfo {author} {\bibfnamefont {Magdalena}\
  \bibnamefont {Zych}}, \bibinfo {author} {\bibfnamefont {Fabio}\ \bibnamefont
  {Costa}}, \bibinfo {author} {\bibfnamefont {Igor}\ \bibnamefont {Pikovski}},
  \ and\ \bibinfo {author} {\bibfnamefont {Caslav}\ \bibnamefont {Brukner}},\
  }\bibfield  {title} {\enquote {\bibinfo {title} {Bell's theorem for temporal
  order},}\ }\href {\doibase 10.1038/s41467-019-11579-x} {\bibfield  {journal}
  {\bibinfo  {journal} {Nat. Commun.}\ }\textbf {\bibinfo {volume} {10}},\
  \bibinfo {pages} {3772} (\bibinfo {year} {2019})}\BibitemShut {NoStop}%
\bibitem [{\citenamefont {Wechs}\ \emph {et~al.}(2019)\citenamefont {Wechs},
  \citenamefont {Abbott},\ and\ \citenamefont {Branciard}}]{Wechs2019}%
  \BibitemOpen
  \bibfield  {author} {\bibinfo {author} {\bibfnamefont {Julian}\ \bibnamefont
  {Wechs}}, \bibinfo {author} {\bibfnamefont {Alastair~A.}\ \bibnamefont
  {Abbott}}, \ and\ \bibinfo {author} {\bibfnamefont {Cyril}\ \bibnamefont
  {Branciard}},\ }\bibfield  {title} {\enquote {\bibinfo {title} {On the
  definition and characterisation of multipartite causal (non)separability},}\
  }\href {\doibase 10.1088/1367-2630/aaf352} {\bibfield  {journal} {\bibinfo
  {journal} {New J. Phys.}\ }\textbf {\bibinfo {volume} {21}},\ \bibinfo
  {pages} {013027} (\bibinfo {year} {2019})}\BibitemShut {NoStop}%
\bibitem [{\citenamefont {Henderson}\ \emph {et~al.}(2020)\citenamefont
  {Henderson}, \citenamefont {Belenchia}, \citenamefont {Castro-Ruiz},
  \citenamefont {Budroni}, \citenamefont {Zych}, \citenamefont {Brukner},\ and\
  \citenamefont {Mann}}]{Henderson2020}%
  \BibitemOpen
  \bibfield  {author} {\bibinfo {author} {\bibfnamefont {Laura~J.}\
  \bibnamefont {Henderson}}, \bibinfo {author} {\bibfnamefont {Alessio}\
  \bibnamefont {Belenchia}}, \bibinfo {author} {\bibfnamefont {Esteban}\
  \bibnamefont {Castro-Ruiz}}, \bibinfo {author} {\bibfnamefont {Costantino}\
  \bibnamefont {Budroni}}, \bibinfo {author} {\bibfnamefont {Magdalena}\
  \bibnamefont {Zych}}, \bibinfo {author} {\bibfnamefont {Caslav}\ \bibnamefont
  {Brukner}}, \ and\ \bibinfo {author} {\bibfnamefont {Robert~B.}\ \bibnamefont
  {Mann}},\ }\bibfield  {title} {\enquote {\bibinfo {title} {Quantum temporal
  superposition: The case of quantum field theory},}\ }\href {\doibase
  10.1103/PhysRevLett.125.131602} {\bibfield  {journal} {\bibinfo  {journal}
  {Phys. Rev. Lett.}\ }\textbf {\bibinfo {volume} {125}},\ \bibinfo {pages}
  {131602} (\bibinfo {year} {2020})}\BibitemShut {NoStop}%
\bibitem [{\citenamefont {Ballentine}(1986)}]{Ballentine1986}%
  \BibitemOpen
  \bibfield  {author} {\bibinfo {author} {\bibfnamefont {L.~E.}\ \bibnamefont
  {Ballentine}},\ }\bibfield  {title} {\enquote {\bibinfo {title} {Probability
  theory in quantum mechanics},}\ }\href {\doibase
  doi:http://dx.doi.org/10.1119/1.14783} {\bibfield  {journal} {\bibinfo
  {journal} {Am. J. Phys.}\ }\textbf {\bibinfo {volume} {54}},\ \bibinfo
  {pages} {883--889} (\bibinfo {year} {1986})}\BibitemShut {NoStop}%
\bibitem [{\citenamefont {Chiara}\ \emph {et~al.}(2018)\citenamefont {Chiara},
  \citenamefont {Giuntini}, \citenamefont {Leporini},\ and\ \citenamefont
  {Sergioli}}]{DallaChiaraQIL}%
  \BibitemOpen
  \bibfield  {author} {\bibinfo {author} {\bibfnamefont {Maria Luisa~Dalla}\
  \bibnamefont {Chiara}}, \bibinfo {author} {\bibfnamefont {Roberto}\
  \bibnamefont {Giuntini}}, \bibinfo {author} {\bibfnamefont {Roberto}\
  \bibnamefont {Leporini}}, \ and\ \bibinfo {author} {\bibfnamefont {Giuseppe}\
  \bibnamefont {Sergioli}},\ }\enquote {\bibinfo {title} {The mathematical
  environment of quantum information},}\ in\ \href {\doibase
  10.1007/978-3-030-04471-8_1} {\emph {\bibinfo {booktitle} {Quantum
  Computation and Logic: How Quantum Computers Have Inspired Logical
  Investigations}}},\ \bibinfo {editor} {edited by\ \bibinfo {editor}
  {\bibfnamefont {Maria~Luisa}\ \bibnamefont {Dalla~Chiara}}, \bibinfo {editor}
  {\bibfnamefont {Roberto}\ \bibnamefont {Giuntini}}, \bibinfo {editor}
  {\bibfnamefont {Roberto}\ \bibnamefont {Leporini}}, \ and\ \bibinfo {editor}
  {\bibfnamefont {Giuseppe}\ \bibnamefont {Sergioli}}}\ (\bibinfo  {publisher}
  {Springer International Publishing},\ \bibinfo {address} {Cham},\ \bibinfo
  {year} {2018})\BibitemShut {NoStop}%
\bibitem [{\citenamefont {Cox}(1961)}]{Cox}%
  \BibitemOpen
  \bibfield  {author} {\bibinfo {author} {\bibfnamefont {R.T.}\ \bibnamefont
  {Cox}},\ }\href@noop {} {\emph {\bibinfo {title} {Algebra of Probable
  Inference}}}\ (\bibinfo  {publisher} {Johns Hopkins University Press},\
  \bibinfo {year} {1961})\BibitemShut {NoStop}%
\bibitem [{\citenamefont {Fine}(1973)}]{Fine}%
  \BibitemOpen
  \bibfield  {author} {\bibinfo {author} {\bibfnamefont {T.L.}\ \bibnamefont
  {Fine}},\ }\href@noop {} {\emph {\bibinfo {title} {Theories of probability:
  an examination of foundations}}}\ (\bibinfo  {publisher} {Academic Press},\
  \bibinfo {year} {1973})\BibitemShut {NoStop}%
\bibitem [{\citenamefont {Jaynes}\ and\ \citenamefont
  {Bretthorst}(2003)}]{Jaynes}%
  \BibitemOpen
  \bibfield  {author} {\bibinfo {author} {\bibfnamefont {E.T.}\ \bibnamefont
  {Jaynes}}\ and\ \bibinfo {author} {\bibfnamefont {G.L.}\ \bibnamefont
  {Bretthorst}},\ }\href@noop {} {\emph {\bibinfo {title} {Probability Theory:
  The Logic of Science}}}\ (\bibinfo  {publisher} {Cambridge University
  Press},\ \bibinfo {year} {2003})\BibitemShut {NoStop}%
\bibitem [{\citenamefont {Griffiths}(1984)}]{Griffiths1984}%
  \BibitemOpen
  \bibfield  {author} {\bibinfo {author} {\bibfnamefont {R.~B.}\ \bibnamefont
  {Griffiths}},\ }\bibfield  {title} {\enquote {\bibinfo {title} {Consistent
  histories and the interpretation of quantum mechanics},}\ }\href@noop {}
  {\bibfield  {journal} {\bibinfo  {journal} {J. Stat. Phys.}\ }\textbf
  {\bibinfo {volume} {36}},\ \bibinfo {pages} {219--272} (\bibinfo {year}
  {1984})}\BibitemShut {NoStop}%
\bibitem [{\citenamefont {Griffiths}(2003)}]{Griffiths}%
  \BibitemOpen
  \bibfield  {author} {\bibinfo {author} {\bibfnamefont {R.~B.}\ \bibnamefont
  {Griffiths}},\ }\href@noop {} {\emph {\bibinfo {title} {Consistent Quantum
  Theory}}}\ (\bibinfo  {publisher} {Cambridge University Press},\ \bibinfo
  {year} {2003})\BibitemShut {NoStop}%
\bibitem [{\citenamefont {Omnès}(1992)}]{Omnes1992}%
  \BibitemOpen
  \bibfield  {author} {\bibinfo {author} {\bibfnamefont {Roland}\ \bibnamefont
  {Omnès}},\ }\bibfield  {title} {\enquote {\bibinfo {title} {Consistent
  interpretations of quantum mechanics},}\ }\href@noop {} {\bibfield  {journal}
  {\bibinfo  {journal} {Rev. Mod. Phys.}\ }\textbf {\bibinfo {volume} {64}},\
  \bibinfo {pages} {339--382} (\bibinfo {year} {1992})}\BibitemShut {NoStop}%
\bibitem [{\citenamefont {Itzykson}\ and\ \citenamefont
  {Zuber}(1985)}]{Itzykson-Zuber}%
  \BibitemOpen
  \bibfield  {author} {\bibinfo {author} {\bibfnamefont {C.}~\bibnamefont
  {Itzykson}}\ and\ \bibinfo {author} {\bibfnamefont {J.-B.}\ \bibnamefont
  {Zuber}},\ }\href@noop {} {\emph {\bibinfo {title} {Quantum Field Theory}}}\
  (\bibinfo  {publisher} {McGraw-Hill Book Co.},\ \bibinfo {address}
  {Singapore},\ \bibinfo {year} {1985})\BibitemShut {NoStop}%
\bibitem [{\citenamefont {Barnett}(2009)}]{Barnett}%
  \BibitemOpen
  \bibfield  {author} {\bibinfo {author} {\bibfnamefont {S.}~\bibnamefont
  {Barnett}},\ }\href@noop {} {\emph {\bibinfo {title} {Quantum Information}}}\
  (\bibinfo  {publisher} {OUP Oxford},\ \bibinfo {year} {2009})\BibitemShut
  {NoStop}%
\bibitem [{\citenamefont {Laloë}(2019)}]{Laloe}%
  \BibitemOpen
  \bibfield  {author} {\bibinfo {author} {\bibfnamefont {Franck}\ \bibnamefont
  {Laloë}},\ }\href {\doibase DOI: 10.1017/9781108569361} {\emph {\bibinfo
  {title} {Do We Really Understand Quantum Mechanics?}}},\ \bibinfo {edition}
  {2nd}\ ed.\ (\bibinfo  {publisher} {Cambridge University Press},\ \bibinfo
  {address} {Cambridge},\ \bibinfo {year} {2019})\BibitemShut {NoStop}%
\bibitem [{\citenamefont {Auletta}\ \emph {et~al.}(2009)\citenamefont
  {Auletta}, \citenamefont {Fortunato},\ and\ \citenamefont
  {Parisi}}]{AulettaQM}%
  \BibitemOpen
  \bibfield  {author} {\bibinfo {author} {\bibfnamefont {G.}~\bibnamefont
  {Auletta}}, \bibinfo {author} {\bibfnamefont {M.}~\bibnamefont {Fortunato}},
  \ and\ \bibinfo {author} {\bibfnamefont {G.}~\bibnamefont {Parisi}},\
  }\href@noop {} {\emph {\bibinfo {title} {Quantum Mechanics}}}\ (\bibinfo
  {publisher} {Cambridge University Press},\ \bibinfo {year}
  {2009})\BibitemShut {NoStop}%
\bibitem [{\citenamefont {Putnam}(1969)}]{Putnam1969}%
  \BibitemOpen
  \bibfield  {author} {\bibinfo {author} {\bibfnamefont {Hilary}\ \bibnamefont
  {Putnam}},\ }\enquote {\bibinfo {title} {Is logic empirical?}}\ in\ \href
  {\doibase 10.1007/978-94-010-3381-7_5} {\emph {\bibinfo {booktitle} {Boston
  Studies in the Philosophy of Science: Proceedings of the Boston Colloquium
  for the Philosophy of Science 1966/1968}}}\ (\bibinfo  {publisher} {Springer
  Netherlands},\ \bibinfo {address} {Dordrecht},\ \bibinfo {year} {1969})\ pp.\
  \bibinfo {pages} {216--241}\BibitemShut {NoStop}%
\bibitem [{\citenamefont {Piron}(1972)}]{Piron1972}%
  \BibitemOpen
  \bibfield  {author} {\bibinfo {author} {\bibfnamefont {C.}~\bibnamefont
  {Piron}},\ }\bibfield  {title} {\enquote {\bibinfo {title} {Survey of general
  quantum physics},}\ }\href {\doibase 10.1007/bf00708413} {\bibfield
  {journal} {\bibinfo  {journal} {Found. Phys.}\ }\textbf {\bibinfo {volume}
  {2}},\ \bibinfo {pages} {287--314} (\bibinfo {year} {1972})}\BibitemShut
  {NoStop}%
\bibitem [{\citenamefont {Auletta}(2001)}]{Auletta}%
  \BibitemOpen
  \bibfield  {author} {\bibinfo {author} {\bibfnamefont {G.}~\bibnamefont
  {Auletta}},\ }\href@noop {} {\emph {\bibinfo {title} {Foundations and
  Interpretation of Quantum Mechanics: In the Light of a Critical-historical
  Analysis of the Problems and of a Synthesis of the Results}}}\ (\bibinfo
  {publisher} {World Scientific},\ \bibinfo {year} {2001})\BibitemShut
  {NoStop}%
\bibitem [{\citenamefont {{Dalla~Chiara}}\ \emph {et~al.}(2004)\citenamefont
  {{Dalla~Chiara}}, \citenamefont {Giuntini},\ and\ \citenamefont
  {Greechie}}]{DallaChiara}%
  \BibitemOpen
  \bibfield  {author} {\bibinfo {author} {\bibfnamefont {M.L.}\ \bibnamefont
  {{Dalla~Chiara}}}, \bibinfo {author} {\bibfnamefont {R.}~\bibnamefont
  {Giuntini}}, \ and\ \bibinfo {author} {\bibfnamefont {R.}~\bibnamefont
  {Greechie}},\ }\href@noop {} {\emph {\bibinfo {title} {Reasoning in Quantum
  Theory: Sharp and Unsharp Quantum Logics}}}\ (\bibinfo  {publisher} {Springer
  Netherlands},\ \bibinfo {year} {2004})\BibitemShut {NoStop}%
\bibitem [{\citenamefont {Scully}\ \emph {et~al.}(1991)\citenamefont {Scully},
  \citenamefont {Englert},\ and\ \citenamefont {Walther}}]{Scully1991}%
  \BibitemOpen
  \bibfield  {author} {\bibinfo {author} {\bibfnamefont {Marian~O.}\
  \bibnamefont {Scully}}, \bibinfo {author} {\bibfnamefont {Berthold-Georg}\
  \bibnamefont {Englert}}, \ and\ \bibinfo {author} {\bibfnamefont {Herbert}\
  \bibnamefont {Walther}},\ }\bibfield  {title} {\enquote {\bibinfo {title}
  {Quantum optical tests of complementarity},}\ }\href@noop {} {\bibfield
  {journal} {\bibinfo  {journal} {Nature}\ }\textbf {\bibinfo {volume} {351}},\
  \bibinfo {pages} {111--116} (\bibinfo {year} {1991})}\BibitemShut {NoStop}%
\bibitem [{\citenamefont {Herzog}\ \emph {et~al.}(1995)\citenamefont {Herzog},
  \citenamefont {Kwiat}, \citenamefont {Weinfurter},\ and\ \citenamefont
  {Zeilinger}}]{Herzog1995}%
  \BibitemOpen
  \bibfield  {author} {\bibinfo {author} {\bibfnamefont {Thomas~J.}\
  \bibnamefont {Herzog}}, \bibinfo {author} {\bibfnamefont {Paul~G.}\
  \bibnamefont {Kwiat}}, \bibinfo {author} {\bibfnamefont {Harald}\
  \bibnamefont {Weinfurter}}, \ and\ \bibinfo {author} {\bibfnamefont {Anton}\
  \bibnamefont {Zeilinger}},\ }\bibfield  {title} {\enquote {\bibinfo {title}
  {Complementarity and the quantum eraser},}\ }\href@noop {} {\bibfield
  {journal} {\bibinfo  {journal} {Phys. Rev. Lett.}\ }\textbf {\bibinfo
  {volume} {75}},\ \bibinfo {pages} {3034--3037} (\bibinfo {year}
  {1995})}\BibitemShut {NoStop}%
\bibitem [{\citenamefont {Kim}\ \emph {et~al.}(2000)\citenamefont {Kim},
  \citenamefont {Yu}, \citenamefont {Kulik}, \citenamefont {Shih},\ and\
  \citenamefont {Scully}}]{Kim2000}%
  \BibitemOpen
  \bibfield  {author} {\bibinfo {author} {\bibfnamefont {Yoon-Ho}\ \bibnamefont
  {Kim}}, \bibinfo {author} {\bibfnamefont {Rong}\ \bibnamefont {Yu}}, \bibinfo
  {author} {\bibfnamefont {Sergei~P.}\ \bibnamefont {Kulik}}, \bibinfo {author}
  {\bibfnamefont {Yanhua}\ \bibnamefont {Shih}}, \ and\ \bibinfo {author}
  {\bibfnamefont {Marlan~O.}\ \bibnamefont {Scully}},\ }\bibfield  {title}
  {\enquote {\bibinfo {title} {Delayed ``choice'' quantum eraser},}\ }\href
  {\doibase 10.1103/PhysRevLett.84.1} {\bibfield  {journal} {\bibinfo
  {journal} {Phys. Rev. Lett.}\ }\textbf {\bibinfo {volume} {84}},\ \bibinfo
  {pages} {1--5} (\bibinfo {year} {2000})}\BibitemShut {NoStop}%
\bibitem [{\citenamefont {Weisz}\ \emph {et~al.}(2014)\citenamefont {Weisz},
  \citenamefont {Choi}, \citenamefont {Sivan}, \citenamefont {Heiblum},
  \citenamefont {Gefen}, \citenamefont {Mahalu},\ and\ \citenamefont
  {Umansky}}]{Weisz2014}%
  \BibitemOpen
  \bibfield  {author} {\bibinfo {author} {\bibfnamefont {E.}~\bibnamefont
  {Weisz}}, \bibinfo {author} {\bibfnamefont {H.~K.}\ \bibnamefont {Choi}},
  \bibinfo {author} {\bibfnamefont {I.}~\bibnamefont {Sivan}}, \bibinfo
  {author} {\bibfnamefont {M.}~\bibnamefont {Heiblum}}, \bibinfo {author}
  {\bibfnamefont {Y.}~\bibnamefont {Gefen}}, \bibinfo {author} {\bibfnamefont
  {D.}~\bibnamefont {Mahalu}}, \ and\ \bibinfo {author} {\bibfnamefont
  {V.}~\bibnamefont {Umansky}},\ }\bibfield  {title} {\enquote {\bibinfo
  {title} {An electronic quantum eraser},}\ }\href {\doibase
  10.1126/science.1248459} {\bibfield  {journal} {\bibinfo  {journal}
  {Science}\ }\textbf {\bibinfo {volume} {344}},\ \bibinfo {pages} {1363--1366}
  (\bibinfo {year} {2014})}\BibitemShut {NoStop}%
\bibitem [{\citenamefont {Ma}\ \emph {et~al.}(2016)\citenamefont {Ma},
  \citenamefont {Kofler},\ and\ \citenamefont {Zeilinger}}]{Ma2016}%
  \BibitemOpen
  \bibfield  {author} {\bibinfo {author} {\bibfnamefont {Xiaosong}\
  \bibnamefont {Ma}}, \bibinfo {author} {\bibfnamefont {Johannes}\ \bibnamefont
  {Kofler}}, \ and\ \bibinfo {author} {\bibfnamefont {Anton}\ \bibnamefont
  {Zeilinger}},\ }\bibfield  {title} {\enquote {\bibinfo {title}
  {Delayed-choice gedanken experiments and their realizations},}\ }\href
  {\doibase 10.1103/RevModPhys.88.015005} {\bibfield  {journal} {\bibinfo
  {journal} {Rev. Mod. Phys.}\ }\textbf {\bibinfo {volume} {88}},\ \bibinfo
  {pages} {015005} (\bibinfo {year} {2016})}\BibitemShut {NoStop}%
\bibitem [{\citenamefont {Zurek}(2003)}]{Zurek2003}%
  \BibitemOpen
  \bibfield  {author} {\bibinfo {author} {\bibfnamefont {Wojciech~Hubert}\
  \bibnamefont {Zurek}},\ }\bibfield  {title} {\enquote {\bibinfo {title}
  {Decoherence, einselection, and the quantum origins of the classical},}\
  }\href {\doibase 10.1103/RevModPhys.75.715} {\bibfield  {journal} {\bibinfo
  {journal} {Rev. Mod. Phys.}\ }\textbf {\bibinfo {volume} {75}},\ \bibinfo
  {pages} {715--775} (\bibinfo {year} {2003})}\BibitemShut {NoStop}%
\bibitem [{\citenamefont {Schlosshauer}(2005)}]{Schlosshauer2005}%
  \BibitemOpen
  \bibfield  {author} {\bibinfo {author} {\bibfnamefont {Maximilian}\
  \bibnamefont {Schlosshauer}},\ }\bibfield  {title} {\enquote {\bibinfo
  {title} {Decoherence, the measurement problem, and interpretations of quantum
  mechanics},}\ }\href {\doibase 10.1103/RevModPhys.76.1267} {\bibfield
  {journal} {\bibinfo  {journal} {Rev. Mod. Phys.}\ }\textbf {\bibinfo {volume}
  {76}},\ \bibinfo {pages} {1267--1305} (\bibinfo {year} {2005})}\BibitemShut
  {NoStop}%
\bibitem [{\citenamefont {Englert}(1996)}]{Englert1996}%
  \BibitemOpen
  \bibfield  {author} {\bibinfo {author} {\bibfnamefont {Berthold-Georg}\
  \bibnamefont {Englert}},\ }\bibfield  {title} {\enquote {\bibinfo {title}
  {Fringe visibility and which-way information: An inequality},}\ }\href
  {\doibase 10.1103/PhysRevLett.77.2154} {\bibfield  {journal} {\bibinfo
  {journal} {Phys. Rev. Lett.}\ }\textbf {\bibinfo {volume} {77}},\ \bibinfo
  {pages} {2154--2157} (\bibinfo {year} {1996})}\BibitemShut {NoStop}%
\bibitem [{\citenamefont {Rubino}\ \emph {et~al.}(2017)\citenamefont {Rubino},
  \citenamefont {Rozema}, \citenamefont {Feix}, \citenamefont {Araújo},
  \citenamefont {Zeuner}, \citenamefont {Procopio}, \citenamefont {Brukner},\
  and\ \citenamefont {Walther}}]{Rubino2017}%
  \BibitemOpen
  \bibfield  {author} {\bibinfo {author} {\bibfnamefont {Giulia}\ \bibnamefont
  {Rubino}}, \bibinfo {author} {\bibfnamefont {Lee~A.}\ \bibnamefont {Rozema}},
  \bibinfo {author} {\bibfnamefont {Adrien}\ \bibnamefont {Feix}}, \bibinfo
  {author} {\bibfnamefont {Mateus}\ \bibnamefont {Araújo}}, \bibinfo {author}
  {\bibfnamefont {Jonas~M.}\ \bibnamefont {Zeuner}}, \bibinfo {author}
  {\bibfnamefont {Lorenzo~M.}\ \bibnamefont {Procopio}}, \bibinfo {author}
  {\bibfnamefont {Caslav}\ \bibnamefont {Brukner}}, \ and\ \bibinfo {author}
  {\bibfnamefont {Philip}\ \bibnamefont {Walther}},\ }\bibfield  {title}
  {\enquote {\bibinfo {title} {Experimental verification of an indefinite
  causal order},}\ }\href {\doibase 10.1126/sciadv.1602589} {\bibfield
  {journal} {\bibinfo  {journal} {Sci. Adv.}\ }\textbf {\bibinfo {volume}
  {3}},\ \bibinfo {pages} {e1602589} (\bibinfo {year} {2017})}\BibitemShut
  {NoStop}%
\end{thebibliography}%




\appendix

\section{Quantum eraser probabilities} \label{app:QE} 
We consider two path detector bases $\ket{p_1}, \ket{p_2}$, e.g. corresponding to $\ket{a^{det}}, \ket{b^{det}}$ states of the Young's slit case, and a final detection $d$.
We consider a perfect which-path measurement  where $\ket{p_j}$ are associated to the system state $\ket{j}$.
The associated operators $K_j = \sum_i \braket{p_j | p_i} \ket{i}\bra{i}$ is then equivalent to the projectors $P_j = \ket{j}\bra{j}$. 
We consider two different orthogonal states $\ket{q_1}, \ket{q_2}$ related by the unitary transformation $\ket{q_j} = V_{ji} \ket{p_i}$.
For each measurement $q_j$, the associated operator 
\begin{equation}
K'_j  =  \sum_{i} \braket{q_j | p_i} \ket{i}\bra{i} = \sum_i V_{ji} \ket{i}\bra{i}.
\end{equation}

For each single measurement $q_j$, we have 
\begin{multline}
\wp(d \land q_j | s )  = tr( K_d^{} K_j'^{'} \rho K_j^{' \dag} K_d^\dag) = \\
 = tr( K_d (V_{j1} P_1 + V_{j2} P_2)\rho_S ( V_{j1}^* P_1 +V_{j2}^* P_2 ) P_d )= \\
 tr(( \sum_i V_{ji} P_i)  \rho ( \sum_{i'}  V_{ji'}^* P_{i'}) P_d ) = \\
  = \sum_{i,i'} V_{ji} V_{ji'}^*\  tr(P_i  \rho_S  P_{i'} P_d )  =  \\
 = \sum_{i} | V_{ji} | ^2\  tr(P_i  \rho_S  P_{i} P_d )  + \sum_{i,i' \ne i} V_{ji} V_{ji'}^*\  tr(P_i  \rho_S  P_{i'} P_d ) .
\end{multline}

When we consider the probability relative to the measurement $(d \land q_1) \lor (d \land q_2)$, we have
\begin{multline}
\wp((d \land q_1) \lor (d \land q_2) | s )  = \\
=   \sum_{j,i} | V_{ji} | ^2\  tr(P_i  \rho_S  P_{i} P_{d} )  + \sum_{ij,,i' \ne i} V_{ji} V_{ji'}^*\  tr(P_i  \rho_S  P_{i'} P_{d} )    = \\
= \sum_{i}  tr(P_i  \rho_S  P_{i} P_{d} )  + \sum_{i,i' \ne i,j} V_{ji} V_{ji'}^*\  tr(P_i  \rho_S  P_{i'} P_{d} )   
\end{multline}
where we used the unitary matrix property $\sum_{j} | V_{ji} | ^2 =1 $.
The second term of the expression is in fact equal to zero because of other property of unitarity of $V$ matrices  $\sum_{j} V_{ji} V_{ji'}^*   = \delta_{i, i'}$ in a sum over $i, i'\ne i$.
Finally we have
\begin{multline}
\wp((d \land q_1) \lor (d \land q_2) | s )  = \sum_{i}  tr(P_i  \rho  P_{i} P_{d} )  = \\
=  \wp(d \land p_1) \lor (d \land p_2) | s ).
\end{multline}
Independently of the choice of the orthogonal and complete base, probability is always the same and equal to a particle-like behavior, even if the single $d \land q_j$ measurement can provoke interference effects.

\end{document}